\documentclass[aps,prd,twocolumn,10pt
,
tightenlines,superscriptaddress,
nofootinbib,showpacs]{revtex4}
\usepackage{amssymb,latexsym}
\usepackage{amsmath,amsbsy,bbm}
\usepackage{epsfig,bm}
\usepackage{graphicx,comment}
\usepackage{hyperref}

\DeclareMathOperator{\str}{str}
\DeclareMathOperator{\Tr}{Tr}

\newcommand\beq{\begin{eqnarray}}
\newcommand\eeq{\end{eqnarray}}

\begin{document}
\def\vp{{\vec{p}}}
\def\vx{{\vec{x}}}
\def\vD{{\vec{D}}}
\def\vB{{\vec{B}}}
\def\vE{{\vec{E}}}
\def\vcE{{\vec{\mathcal{E}}}}
\def\vA{{\vec{A}}}
\def\vDel{{\vec{\nabla}}}
\def\vzero{{\vec{0}}}

\def\a{{\alpha}}
\def\b{{\beta}}
\def\d{{\delta}}
\def\D{{\Delta}}
\def\X{{\Xi}}
\def\e{{\varepsilon}}
\def\g{{\gamma}}
\def\G{{\Gamma}}
\def\k{{\kappa}}
\def\l{{\lambda}}
\def\L{{\Lambda}}
\def\m{{\mu}}
\def\n{{\nu}}
\def\o{{\omega}}
\def\O{{\Omega}}
\def\S{{\Sigma}}
\def\s{{\sigma}}
\def\th{{\theta}}

\def\ol#1{{\overline{#1}}}
\def\sumint{\sum \hskip-1.35em \int_{ \hskip-0.25em \underset{k}{\phantom{a}} } \, \, \,}

\def\Dslash{D\hskip-0.65em /}
\def\diag{\text{diag}}

\def\cE{{\mathcal E}}
\def\cF{{\mathcal F}}
\def\cG{{\mathcal G}}
\def\cS{{\mathcal S}}
\def\cC{{\mathcal C}}
\def\cB{{\mathcal B}}
\def\cU{{\mathcal U}}
\def\cT{{\mathcal T}}
\def\cQ{{\mathcal Q}}
\def\cL{{\mathcal L}}
\def\cO{{\mathcal O}}
\def\cA{{\mathcal A}}
\def\cQ{{\mathcal Q}}
\def\cR{{\mathcal R}}
\def\cs{{\mathfrak s}}
\def\cH{{\mathcal H}}
\def\cW{{\mathcal W}}
\def\cM{{\mathcal M}}
\def\cD{{\mathcal D}}
\def\cN{{\mathcal N}}
\def\cP{{\mathcal P}}
\def\cK{{\mathcal K}}

\def\eqref#1{{(\ref{#1})}}

\preprint{RBRC-1083}

\title{Finite Volume Effects on the Extraction of Form Factors at Zero Momentum}

\affiliation{
Department of Physics,
        The City College of New York,  
        New York, NY 10031, USA}
\author{Brian~C.~Tiburzi}
\email[]{btiburzi@ccny.cuny.edu}
\affiliation{
Department of Physics,
        The City College of New York,  
        New York, NY 10031, USA}
\affiliation{
Graduate School and University Center,
        The City University of New York,
        New York, NY 10016, USA}
\affiliation{
RIKEN BNL Research Center, 
        Brookhaven National Laboratory, 
        Upton, NY 11973, USA}
\date{\today}


\begin{abstract}
Hadronic matrix elements that depend on momentum are required for numerous phenomenological applications. 
Probing the low-momentum regime is often problematic for lattice QCD computations on account of the restriction to periodic momentum modes. 
Recently a novel method has been proposed to compute matrix elements at zero momentum, 
for which straightforward evaluation of the matrix elements would otherwise yield a vanishing result. 
We clarify an assumption underlying this method, 
and thereby establish the theoretical framework required to address the associated finite volume effects. 
Using the pion electromagnetic form factor as an example, 
we show how the charge radius and two higher moments can be calculated at zero momentum transfer, 
and determine the corresponding finite volume effects. 
These computations are performed using chiral perturbation theory to account for  
modified infrared physics, 
and can be generalized to ascertain finite volume effects for other hadronic matrix elements extracted at 
zero momentum.  
\end{abstract}
\maketitle

\section{Introduction} %

Experimental measurement of the Lamb shift and hyperfine splitting in muonic hydrogen has lead to a determination of the proton charge radius to unprecedented precision%
~\cite{Pohl:2010zza,Antognini:1900ns}.
The extracted value lies 
$7 \, \sigma$
away from the CODATA recommended value for the charge radius%
~\cite{Mohr:2012tt}, 
which is determined from the world's data on electron-proton scattering as well as the spectrum of electronic hydrogen. 
In the wake of such a surprising determination from muonic hydrogen, 
the proton size puzzle has attracted a considerable amount of attention, 
see~\cite{Pohl:2013yb} for a review. 
Given the tremendous advances in lattice gauge theory techniques, 
insight into the proton size puzzle might be garnered from first principles lattice QCD computations. 
For an overview of the current status of nucleon structure computations using lattice QCD, 
see%
~\cite{Syritsyn:2014saa}.

There are many issues confronting the determination of charge radii from lattice QCD.  
In this work, 
we focus on just one aspect of the problem, 
namely the limitation to lattice-quantized momentum transfer. 
This limitation emerges on account of the periodic boundary conditions satisfied by quark fields.   
Many phenomenological applications require knowledge of the momentum-transfer dependence of hadronic matrix elements, 
and this presents a challenge for lattice QCD computations given the coarse-grained sampling of momentum transfer possible with periodic momentum modes. 
Radii, 
for example, 
depend on the slope of form factors evaluated at vanishing momentum transfer. 
As such, 
these quantities are not directly accessible with conventional lattice QCD methods, 
and one often models the momentum-transfer dependence of the matrix element to extract the desired quantity. 
The theoretical situation essentially parallels the experimental problem of extracting radii using form factor data. 
It is thus desirable to remove this source of uncertainty from lattice QCD calculations.

One way to overcome the restriction to lattice-quantized momenta is to impose twisted boundary conditions on the quark fields%
~\cite{Bedaque:2004kc,deDivitiis:2004kq,Sachrajda:2004mi}. 
Indeed for the pion form factor, 
for example, 
twisted boundary conditions have been utilized to better access the pion charge radius, 
see%
~\cite{Flynn:2005in,Guadagnoli:2005be,Boyle:2007wg,Boyle:2008yd,Frezzotti:2008dr}.
Due to computational restrictions, 
however, 
the method of twisted boundary conditions is currently only practicable in a partially twisted scenario%
~\cite{Sachrajda:2004mi,Bedaque:2004ax}, 
i.e.~where the valence quarks are subject to twisted boundary conditions, while sea quarks remain periodic. 
Computationally this means that quark propagators are determined on gauge configurations that are not modified as the twist parameters vary. 
One can view the partially twisted scenario as that of a mixed action%
~\cite{Bar:2002nr}, 
where the valence and sea quarks differ only by their respective boundary conditions.
Unfortunately in this scenario,  
lattice results obtained at different values of twist angles are correlated. 
When such correlations are properly accounted for in fitting the momentum dependence of 
matrix elements, 
there is no guarantee that statistically independent information about the matrix element will be obtained as the twist  varies. 
As a result, 
partially twisted boundary conditions often give one confidence about modeling the momentum dependence of matrix elements, 
however, 
they do not necessarily reduce the statistical uncertainty in the extraction of phenomenologically interesting parameters.

In order to remove uncertainty associated with the momentum dependence of matrix elements,  
a novel method has been recently proposed%
~\cite{deDivitiis:2012vs}.
The method hinges on writing down a Taylor series expansion of lattice correlation functions in terms of the external momenta. 
Instead of computing the momentum dependence of matrix elements, 
one computes the Taylor series coefficients directly at zero momentum. 
The method provides a practical lattice definition of momentum-dependent quantities
possessing the correct infinite volume
limit. 
Because the method is proposed to overcome what is essentially a restriction to finite volume, 
we investigate the associated finite volume effects. 
Given the motivation to reduce uncertainties in extracting the proton charge radius, 
we additionally extend the zero momentum method to charge radii;
however, 
for simplicity, 
we focus on the pion charge radius.

Our investigation is organized as follows. 
We begin in 
Sec.~\ref{s:ZME}
with a simple observation about the nature of the momentum expansion on a lattice of fixed size. 
To derive a Taylor series expansion of correlation functions, 
one requires that the momentum-carrying quark is subject to a twisted boundary condition, 
with the Taylor coefficients arising from differentiation with respect to twist angle, 
and subsequent evaluation at vanishing twist. 
By deriving the method on a lattice of fixed size, 
we are able to develop a framework to address finite volume corrections. 
Next in 
Sec.~\ref{s:AZMAZMT}, 
we consider the electromagnetic form factor of the pion, 
and present correlator derivatives that can cleanly isolate the pion charge radius at zero momentum. 
Two higher moments of the electric charge distribution are also considered. 
A brief discussion is included on the absence of power-law divergent contributions at finite lattice spacing. 
The determination of finite volume corrections to the method at zero momentum is taken up in 
Sec.~\ref{s:FVC}.
Here we formulate partially twisted chiral perturbation theory for twisted initial- and final-state quarks, 
with periodic spectator and sea quarks. 
This enables us to compute the finite volume current matrix element of the pion in an arbitrary frame. 
Subsequent differentiation of this result yields the finite volume effect on the extraction of the charge radius and two higher moments using the zero momentum method. 
A technical detail related to computing momentum derivatives of finite volume mode sums is discussed in 
Appendix~\ref{s:A}. 
Finally a brief summary in
Sec.~\ref{s:S}
concludes our work.

\section{Zero Momentum Expansion} %
\label{s:ZME}

The method of obtaining hadronic form factors at vanishing momentum%
~\cite{deDivitiis:2012vs}
hinges on an expansion of correlation functions in powers of the momenta. 
A generic correlation function depending on the three-momentum 
$\vec{p}\,$
and Euclidean time 
$x_4$
is written as 
\begin{eqnarray}
C(\vec{p} \, , x_4 )
&=&
\sum_{\vec{x}} e^{ - i \vec{p} \cdot \vec{x}}
\int D\cU \, \cP[\cU] C[x, \cU] 
\notag \\
&\equiv&
\int D\cU \, \cP[\cU] C[\vec{p} \, , x_4 , \cU]
,\end{eqnarray}
where 
$\cU$
are gauge links, 
and 
$\cP[\cU]$
their corresponding probabilistic weight in the functional integration. 
In a fixed gauge background, 
one can 
\emph{formally}
write a momentum expansion of the correlation function having the form
\begin{eqnarray}
C[\vec{p}\, , x_4 , \cU] 
&=& 
C^{(0)} [x_4, \cU]
+ 
p_i C^{(1)}_i [x_4, \cU]
\notag \\
&& \phantom{sp}
+ 
\frac{1}{2} p_i p_j 
C^{(2)}_{ij} [x_4, \cU] 
+ 
\ldots
\label{eq:Taylor}
,\end{eqnarray}
where the coefficients are to be identified as those of a Taylor series expansion about vanishing momentum
\begin{equation}
C^{(n)}_{i\cdots} [x_4, \cU] 
= 
\frac{\partial^n C[\vec{p} \, , x_4 ,\cU]}{\partial p_i \cdots} \Bigg|_{\vec{p} = \vec{0}}
.\end{equation} 
The heart of the method is to compute such coefficients directly by similarly expanding quark propagators 
and vertices in powers of the momenta. 
These expansions yield expressions for the coefficients as modified correlation functions depending only on periodic propagators evaluated at vanishing momentum. 
In this way, 
one circumvents the need to evaluate the correlation function 
$C(\vec{p} \,, x_4 )$
as a function of 
$\vec{p}$,
and subsequently perform an extrapolation to obtain the coefficient
$\partial C(\vec{p}\,, x_4 ) /\partial p_i  \big|_{\vec{p} = \vec{0}}$, 
for example.

In considering the Taylor series expansion in 
Eq.~\eqref{eq:Taylor}, 
we note that the expansion necessarily has the correct infinite volume limit, 
by construction. 
In this limit, 
the momenta become continuous variables, 
and differentiation leads at once to the Taylor series expansion. 
In this work, 
we concern ourselves with finite volume corrections that arise from employing the method. 
In finite volume, 
by contrast, 
the expansion in 
Eq.~\eqref{eq:Taylor}
can only be justified by using a twisted boundary condition on the active quark, 
and the natural question becomes how to assess the associated finite volume effects.

To see that the expansion is only justified for a twisted boundary condition, 
we exemplify the case of a two-particle correlation function in a free scalar field theory. 
Because this theory describes only non-interacting particles, 
we are to set all gauge links to unity, 
and omit the functional integration over 
$\cU$. 
To keep the discussion as simple as possible, 
we furthermore take the continuum limit. 
Consider a compact space of length
$L$
in each of the three spatial directions.  
The periodic single-particle propagator, 
$S(x',x)$,
is given by
\begin{equation}
S(x',x)
=
\frac{1}{L^3}
\sum_{\vec{n}}
e^{ 2 \pi i \vec{n} \cdot (\vec{x}' - \vec{x}) / L} 
\frac{ e^{ - E \left(\frac{2 \pi \vec{n}}{L}  \right) | x'_4 - x_4|}}{2 E \left(\frac{2 \pi \vec{n}}{L}  \right)}
,\end{equation}
where the energy 
$E \left(\vec{p}\, \right)$ 
satisfies 
$E \left(\vec{p}\, \right)^2 = \vec{p}\, {}^2 + m^2$, 
with
$E \left(\vec{p}\, \right) > 0$.
By virtue of the periodic boundary conditions satisfied by the scalar field, 
the spatial momenta are quantized in the form
 \begin{equation} \label{eq:p}
\vec{p} = \frac{2 \pi}{L} \vec{n}
,\end{equation} 
where
$\vec{n}\, $
is a triplet of integers;
and, 
in the propagator, 
we have written
$\sum_{\vec{n}}$
as a shorthand notation for the summation over all momentum mode numbers,
$\sum_{n_1 = -\infty}^\infty  \sum_{n_2 = - \infty}^\infty  \sum_{n_3 = - \infty}^\infty$.

The two-particle correlation function with total three-momentum 
$\vec{p}\,$
is defined by 
\begin{equation}
C(\vec{p}\, , x_4)
=
\int_0^L
d \vec{x} \,
e^{ - i \vec{p} \cdot \vec{x}} 
S(x,0) S(0,x)
,\end{equation}
where the integral 
$\int_0^L d \vec{x}$
is an abbreviation for the volume integration
$\int_0^L dx_1 \int_0^L dx_2 \int_0^L dx_3$. 
In the non-interacting scalar theory, 
it is trivial to evaluate the two-particle correlation function. 
When 
$\vec{p} \, $
is a periodic momentum mode satisfying the quantization condition in 
Eq.~\eqref{eq:p}, 
we arrive at
\begin{equation}
C\left(
\frac{2 \pi \vec{n}}{L}  , x_4
\right)
=
\frac{1}{L^3}
\sum_{\vec{n}'}
\frac{
e^{ - E \left(\frac{2 \pi \vec{n}\,' }{L} \right)  | x_4| }
e^{ - E \left(\frac{2 \pi [\vec{n} \, ' + \vec{n}]}{L}  \right)  | x_4| }
}
{2 E \left(\frac{2 \pi \vec{n} \, '}{L}  \right)  2 E \left(\frac{2 \pi  [\vec{n} \, ' + \vec{n}]}{L} \right) }
.\end{equation}
Because all fields are periodic, 
the only parameter that distinguishes the external momentum, 
$\vec{p} = \frac{2 \pi}{L} \vec{n} $,  
from the internal momentum, 
$\vec{p}  \, ' = \frac{2 \pi}{L} \vec{n}  \, '$,  
is the vector of integers 
$\vec{n}$. 
A Taylor series expansion of this two-particle correlation function in powers of 
$\vec{n}$,  
which is required to arrive at Eq.~\eqref{eq:Taylor}, 
cannot be mathematically justified.
\footnote{
There is a way to approximate derivatives with respect to a quantized momentum. 
Such approximations, 
however, 
hinge on taking large mode numbers, 
and employ differences between such successive 
modes.  
For example, 
the relative change between successive modes,
$\Delta p_i / p_i = 1 / n_i$,
becomes infinitesimally small as 
$n_i$
becomes large. 
In order for large mode numbers to correspond to small momenta, 
however, 
one requires prohibitively large lattices, 
see%
~\cite{Tiburzi:2007ep}. 
The successive difference approach, 
moreover, 
is not related to the method proposed in%
~\cite{deDivitiis:2012vs}.
}

\begin{figure}
\includegraphics[width=0.3\textwidth]{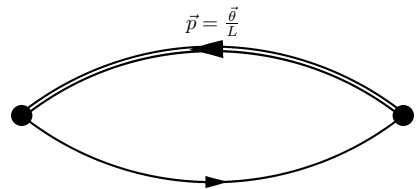}%
\caption{
Graphical depiction of the modified correlation function
$\cC$. 
The single line represents the spectator particle, 
while the double line represents the particle carrying momentum. 
The twist angles
$\vec{\theta}$
allow the momentum to be varied continuously, 
and justify the use of a Taylor series expansion in a fixed volume. 
}
\label{f:exmpl}
\end{figure}

To derive the momentum expansion rigorously in the context of this simple example,
let us instead begin with a scalar field satisfying twisted boundary conditions. 
The corresponding single-particle propagator, 
$S_{\vec{\theta} \,} (x',x)$,
consequently obeys
\begin{equation}
S_{\vec{\theta} \,} (x' + L \hat{e}_j,x) = e^{i \theta_j} S_{\vec{\theta}\,} (x',x)
,\end{equation} 
where 
$\hat{e}_j$ 
is a unit vector in the 
$j$-th 
spatial direction, 
and each of the components of 
$\vec{\theta} \,$
can be varied continuously. 
This propagator is then used for one of the particles in a modified definition for the two-particle correlation function
\begin{equation}
\cC
\left(
\frac{\vec{\theta} }{L}, 
x_4
\right)
=
\int_0^L
d \vec{x} \,
e^{ - i \vec{\theta} \cdot \vec{x}  / L }
S_{\vec{\theta} \, }
(x,0) 
S
(0,x)
\label{eq:mod}
,\end{equation}
and this definition reduces to the previous case when each component of 
$\vec{\theta}\, $
is an integer multiple of 
$2 \pi$. 
A graphical depiction of this simple correlation function is shown in 
Fig.~\ref{f:exmpl}. 
To compute the two-particle correlation function, 
we associate the phase factor with the active particle
by defining the propagator
\begin{equation}
S(x', x | \vec{\theta} \,)
= 
e^{ - i \vec{\theta} \cdot (\vec{x} \, ' - \vec{x}\, )  / L }
S_{\vec{\theta} \, } (x',x)
,\end{equation}
where this propagator is strictly periodic, 
but is that of a scalar field coupled to the uniform gauge potential
$\vec{A} = \frac{\vec{\theta}}{L}$.
From the explicit form of this propagator, 
namely
\begin{equation}
S(x',x | \vec{\theta}\, ) 
= 
\frac{1}{L^3}
\sum_{\vec{n}}
e^{ 2 \pi i \vec{n} \cdot (\vec{x}' - \vec{x}) / L} 
\frac{ 
e^{ - E \left(\frac{ 2 \pi \vec{n}  + \vec{\theta}}{L} \right) | x'_4 - x_4|}
}{2 E \left(\frac{ 2 \pi \vec{n}  + \vec{\theta}}{L} \right) }
,\end{equation} 
we see that the modified two-particle correlation function in 
Eq.~\eqref{eq:mod}
is given by
\begin{equation}
\cC(\vec{p}\, , x_4)
=
\frac{1}{L^3}
\sum_{\vec{n}}
\frac{
e^{ - E \left(\frac{ 2 \pi \vec{n}}{L} \right) | x_4| }
e^{ - E \left(\frac{ 2 \pi \vec{n}}{L} + \vec{p} \, \right)  | x_4| }
}
{2 E \left(\frac{ 2 \pi \vec{n}}{L} \right) \, 2 E \left(\frac{ 2 \pi \vec{n}}{L} + \vec{p} \,  \right) }
,\end{equation} 
where the external momentum satisfies the relation
\begin{equation}
\vec{p} = \frac{\vec{\theta}}{L}
.\end{equation}  
Because the twist angles can be continuously varied, 
one has the trivial equivalence 
$p_i \frac{\partial}{\partial p_i} = \theta_i \frac{\partial}{\partial \theta_i}$
that justifies the Taylor series expansion of this correlation function, 
as in 
Eq.~\eqref{eq:Taylor}. 

%
%
%
%
%
%

We have demonstrated that the Taylor series expansion of correlation functions, 
Eq.~\eqref{eq:Taylor}, 
can be mathematically justified on a finite lattice of fixed size by imposing a twisted boundary condition on the active quark, 
differentiating with respect to the twist angle, 
and finally evaluating at vanishing twist. 
Because the active quark is singled out in this procedure, 
the underlying theoretical framework is strictly speaking not unitary, 
however, 
any violations of unitarity can only appear as finite volume corrections. 
In employing partially twisted chiral perturbation theory below, 
however,
we do not encounter any unitarity violations. 
An additional complication for a general correlation function, 
such as a current matrix element, 
is that there can be numerous active quarks, 
and one must sum over all possible momentum carriers. 
Lastly 
we must stress that the procedure employed by%
~\cite{deDivitiis:2012vs}
is not altered in any way. 
One still computes modified lattice correlation functions corresponding to the Taylor coefficients with periodic propagators at vanishing momentum. 
To address finite volume effects, 
by contrast, 
we require the underlying theoretical framework that justifies the use of the Taylor expansion in 
Eq.~\eqref{eq:Taylor}
on a fixed-size lattice. 
This framework is that of partially twisted boundary conditions imposed on the active quarks, 
which we develop below.

\section{At Zero Momentum and Zero Momentum Transfer} \label{s:AZMAZMT} %

We focus on a particular extension of the zero momentum method, 
namely to the pion's electromagnetic form factor, 
$F(q^2)$.
In infinite volume, 
this form factor is defined through the current matrix element between pion states
\begin{equation} \label{eq:MEL}
\langle \pi(\vec{p} \, ') | J_\mu | \pi ( \vec{p} \,) \rangle
= 
e
(p' + p)_\mu 
F(q^2)
,\end{equation}
where 
$q$
is the momentum transfer between the initial and final states, 
$q_\mu = ( p' - p)_\mu$, 
and 
$e$
is the unit of electric charge. 
The emergence of only one form factor is a consequence of the Lorentz covariance of the matrix element, 
and current conservation. 
Due to properties under charge conjugation, 
we have the relations,
$F_{\pi^+} (q^2) = - F_{\pi^-} (q^2)$, 
and
$F_{\pi^0} (q^2) = 0$. 
There is thus only one pion form factor to consider, 
and we take it to be that of the positively charged pion. 
In the forward limit, 
which is specified by
$p'_\mu = p_\mu$, 
the value of the form factor is determined from the Ward identity as the electric charge of the pion, 
namely
$F(0) = 1$.
The behavior of the form factor away from the forward limit reflects the charge distribution within the pion.
Moments of the charge distribution%
\footnote{
Strictly speaking, 
the Fourier transform of the pion electromagnetic form factor is not a charge density,
however, 
the form factor can be written as the Fourier transform of the transverse distribution of charge in the infinite momentum frame, 
see%
~\cite{Burkardt:2000za,Burkardt:2002hr}.  
}
appear in the momentum-transfer expansion of the form factor. 
We write this expansion in the form
\begin{equation}
F(q^2)
=
1
+
\sum_{n=1}^\infty
\frac{(-q^2)^n}{(2n)! (2n+1)!!}
<r^{2n}> 
\label{eq:moments}
,\end{equation}
where 
$<r^2>$
is the mean-square radius, 
and so on. 
We will focus on just the lowest few moments. 
With vanishing momentum transfer, 
$q_\mu = 0$, 
none of the moments of the pion's charge distribution appear the current matrix element in 
Eq.~\eqref{eq:MEL}.

The pion form factor is accessible through the calculation of three-point correlation functions using lattice QCD. 
We will take the limit of strong isospin, 
and accordingly work with a single light quark field, 
denoted by 
$\psi$. 
Taking a simple point-like pion interpolating field, 
$\pi(x) \sim \ol \psi i  \gamma_5 \psi (x)$, 
we have the vector current three-point correlation function 
\begin{eqnarray}
C_\mu
(\vec{p} \, ', \vec{p} \, | x_4, y_4 )
&=&
-
\sum_{\vec{x},\vec{y}}
e^{- i \vec{p} \, ' \cdot (\vec{x} - \vec{y})} 
e^{- i \vec{p} \cdot \vec{y}}
\notag \\
&&
\times
\left\langle 
\ol \psi \gamma_5 \psi(x) \,
\ol \psi  
\gamma_\mu 
\psi (y) \,
\ol \psi \gamma_5 \psi(0)
\right\rangle
, \quad  \quad
\end{eqnarray}
where
$x_4$
is the Euclidean time separation between the source and sink, 
and 
$y_4$
is the current insertion time. 
Both of these times are assumed positive throughout. 
The brackets refer to the stochastic average over gauge configurations. 
Notice that due to strong isospin invariance, 
the matrix element of the quark electromagnetic current, 
$J_\mu = \frac{2}{3} \ol u \gamma_\mu u - \frac{1}{3} \ol d \gamma_\mu d$,
can be computed from a single matrix element of the vector current,
$V_\mu = \ol \psi \gamma_\mu \psi$, 
or its point-split form.
\footnote{ 
For our particular application of the pion electromagnetic form factor, 
we utilize only the temporal component of the current. 
Expansion of the three-point correlation function in powers of spatial momenta does not require us 
to consider the momentum expansion of the temporal component of the point-split vector current vertex. 
The spatial components of the three-point correlation function, 
on the other hand, 
additionally require expanding the vector current vertex when the point-split current is utilized.  
}

\begin{figure}
\includegraphics[width=0.3\textwidth]{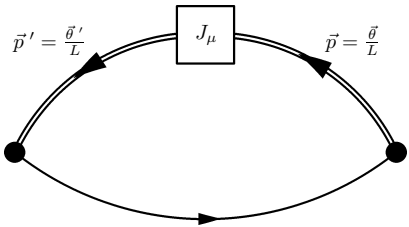}%
\caption{
Quark-level contractions of the modified three-point function, 
$\cC_\mu$. 
The single line represents the spectator quark, 
while the double lines represent the twisted initial- and final-state quarks. 
The twist angles 
$\vec{\theta}$
and
$\vec{\theta} \, '$
are devices that are required to derive the momentum expansion of the current 
correlation function on a lattice of fixed size. 
They are set to zero after differentiation. 
}
\label{f:stuff}
\end{figure}

Performing the quark contractions, 
the three-point correlation function becomes
\begin{eqnarray}
C_\mu
(\vec{p} \, ', \vec{p} \, | x_4, y_4 )
&=&
\sum_{\vec{x},\vec{y}}
e^{- i \vec{p} \, ' \cdot (\vec{x} - \vec{y})} 
e^{- i \vec{p} \cdot \vec{y}}
\notag \\
&&
\times
\left\langle 
\Tr
\left[ \gamma_5 S(x,y) \gamma_\mu S(y,0) \gamma_5 S(0,x) \right]
\right\rangle
,
\notag \\ 
\end{eqnarray}
where 
$S(x',x)$
is used to denote the quark propagators in a background gauge configuration. 
On account of isospin symmetry and the charge conjugation properties of pions, 
there are no self-contractions of the current to evaluate%
~\cite{Draper:1988bp,Bunton:2006va}.  
To utilize a Taylor series expansion in both initial- and final-state momenta, 
we replace the above correlation function with a modified one in which the active quarks are coupled to different uniform gauge fields
\begin{eqnarray}
\cC_\mu
(\vec{p} \, ', \vec{p} \, | x_4, y_4 )
&=&
\sum_{\vec{x},\vec{y}}
\left\langle 
\Tr
\left[ \gamma_5 S(x,y | \vec{\theta} \, ') 
\right.
\right.
\notag \\
&& \times
\left. \left.
\gamma_\mu 
S(y,0 | \vec{\theta} \, ) \gamma_5 S(0,x) \right]
\right\rangle
\label{eq:master}
.\end{eqnarray}
Now the initial- and final-state momenta are continuous parameters that satisfy 
$\vec{p} = \frac{\vec{\theta}}{L}$
and
$\vec{p} \, ' = \frac{\vec{\theta}\,'}{L}$.
The correlation function distinguishes between an initial quark, a final quark, and a spectator quark, 
see Fig.~\ref{f:stuff}.

%
%
%
%
%
%

Assuming a long Euclidean time separation between source and sink, 
as well as from operator insertion to sink, 
we have the following expected behavior for the temporal component of the vector-current correlation function
\begin{eqnarray}
\cC_4
(\vec{p} \, ', \vec{p} \, | x_4, y_4 )
&=&
i [  E \left( \vec{p} \, '\right) + E \left( \vec{p}\,\right) ]
F(q^2)
\notag \\
&& 
\times
|Z|^2
\frac{
e^{ - E\left(\vec{p} \, ' \right) (x_4 - y_4)}
e^{- E\left(\vec{p} \, \right) y_4}
}
{
2 E\left( \vec{p} \, '\right) 
2 E\left( \vec{p} \, \right)
}
\label{eq:VCCF}
,\quad \end{eqnarray}
where we have dropped exponentially suppressed contributions in the Euclidean time separations. 
Such contributions arise from excited states. 
The factor 
$Z$
is the unknown overlap between the interpolating field and the ground-state pion. 
Because the initial- and final-state momenta are continuous, 
we can perform a Taylor series expansion to obtain the charge radius and higher moments of the charge distribution. 
To aid in isolating these desired quantities, 
we observe that the Euclidean momentum transfer takes the form
\begin{equation}
q^2 
= 
2 
\left[
E\left( \vec{p} \, ' \right)
E\left( \vec{p} \, \right)
- 
m_\pi^2
- 
\vec{p} \, ' \cdot \vec{p} \, 
\right]
.\end{equation}
As a consequence, 
the charge radius can be determined from the ratio
\begin{eqnarray}
< r^2 > 
=
\frac{3}{\cC_4 (\vec{0}, \vec{0})}
\frac{
\partial^2 \cC_4 ( \vec{p} \, ', \vec{p} \, )}{\partial p'_1 \partial p_1} \Big|_{\vec{p}\,' = \vec{p} = \vec{0}}
\label{eq:r2}
,\end{eqnarray}
up to exponentially suppressed contributions in the Euclidean time separations. 
Generalizations of this formula allow access to two higher moments of the charge distribution, 
namely
\begin{eqnarray}
<r^4>
&=&
\frac{45}{\cC_4 (\vec{0}, \vec{0})}
\frac{
\partial^4 \cC_4 ( \vec{p} \, ', \vec{p} \, )}{\partial p'_1 \partial p_1 \partial p'_2 \partial p_2} 
\Big|_{\vec{p}\,' = \vec{p} = \vec{0}}
\label{eq:r4}
,
\\
<r^6>
&=&
\frac{1575}{\cC_4 (\vec{0}, \vec{0})}
\frac{
\partial^6 \cC_4 ( \vec{p} \, ', \vec{p} \, )}{\partial p'_1 \partial p_1 \partial p'_2 \partial p_2 \partial p'_3 \partial p_3} 
\Big|_{\vec{p}\,' = \vec{p} = \vec{0}}
\label{eq:r6}
.\quad \end{eqnarray}
Beyond these quantities, 
we have been unable to find simple ratios that isolate any higher moments without the introduction of power-law 
Euclidean time dependence which would contaminate the signal. 
In fact, 
any momentum derivatives of rest frame 
($\vec{p} = \vec{0}$ and $\vec{p}\,' = \vec{q}$)
and Breit frame 
($\vec{p} = - \frac{1}{2} \vec{q}$ and $\vec{p}\,' = \frac{1}{2} \vec{q}$)
three-point correlation functions suffer the same sickness. 
For this reason, 
we consider the three-point function in an arbitrary frame, 
and have found ratios that cleanly isolate the desired coefficients without introducing power-law contamination.  
Notice that for notational ease, 
we have suppressed the Euclidean time dependence of the three-point correlation functions. 
This time dependence will drop out of the correlator ratios, 
provided the ground-state pion saturates the three-point function.

The various momentum derivatives required of the temporal component of the vector three-point function 
in 
Eqs.~\eqref{eq:r2}--\eqref{eq:r6} 
can be expressed in terms of derivatives of quark propagators.
This remains true with a point-split current due to our utilization of the temporal component of the current, 
i.e.~there are no derivatives of the current vertex required for this particular application. 
The derivatives of quark propagators can be expressed in terms of derivatives of the lattice Dirac operator in a uniform 
$U(1)$
gauge field. 
Using the following abstract notation for quark propagators and the Dirac operator 
\begin{eqnarray}
S(x',x| \vec{\theta}\,) 
&=& 
\langle x' | S( \vec{\theta} \, ) | x \rangle
,\notag \\
D(x',x| \vec{\theta}\,)
&=&
\langle x' | D(\vec{\theta} \, ) | x \rangle
,\end{eqnarray}
with 
$D(\vec{\theta} \, ) S(\vec{\theta} \, ) = 1$, 
we have
\begin{equation}
\frac{\partial S(\vec{\theta} \,)}{\partial \theta_i}
=
- S(\vec{\theta} \, ) 
\frac{\partial 
D (\vec{\theta} \, )}{\partial \theta_i}
S( \vec{\theta} \,)
.\end{equation}
For the clover-improved Wilson action, 
the lattice Dirac operator for a quark in a uniform gauge potential has the form
\begin{eqnarray}
D(x',x | \vec{\theta} \, )
&=&
\frac{1}{2}
\sum_{j=1}^3
\Big[
\delta_{x'+j, x}
(\gamma_j  - 1)
e^{  \frac{i \theta_j }{N}}
\cU_j (x')
\notag \\
&& \phantom{spa}
-
\delta_{x',x+j}
(\gamma_j + 1)
e^{- \frac{i \theta_j}{N}}
\cU^\dagger_j (x)
\Big]
+ 
\cdots,
\notag \\
\end{eqnarray}
where we have written only the 
$\vec{\theta}$-dependent terms explicitly, 
and
$N$
is the number of lattice sites in each spatial direction, 
which is just the length 
$L$ 
in lattice units. 
The required momentum derivative of the clover action is simply
\begin{eqnarray}
\frac{\partial D(x',x | \vec{\theta} \,)}{\partial p_i}
&=&
\frac{i}{2}
\Big[
\delta_{x'+i, x}
(\gamma_i  - 1)
e^{\frac{i \theta_i }{ N}}
\cU_i (x')
\notag \\
&&
\phantom{s}
+
\delta_{x',x+i}
(\gamma_i + 1)
e^{-\frac{ i \theta_i}{ N}}
\cU^\dagger_i (x)
\Big]
.\quad
\end{eqnarray}
Valuable simplifications arise from noticing that mixed partials of the lattice Dirac operator must vanish
\begin{equation}
\frac{\partial^2}{\partial p_i \partial p_{j \neq i}}   D(\vec{\theta} \, ) = 0
.\end{equation}
Due to the form of  
Eqs.~\eqref{eq:r2}--\eqref{eq:r6}, 
moreover,
only first partial derivatives are required to determine the charge radius and the two higher moments.

\begin{figure}
\includegraphics[width=0.3\textwidth]{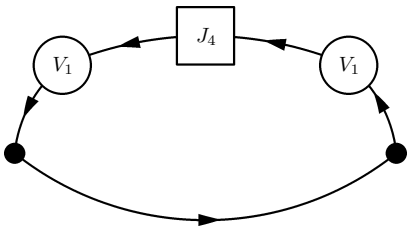}%
\caption{
Quark-level contractions required to obtain the pion charge radius at vanishing momentum. 
The solid circles represent the pion source and sink, 
the square represents the insertion of the external current, 
while the the open circles are the momentum derivative vertices. 
Two higher moments of the charge distribution can be obtained by further insertions of such vertices on the initial- and final-state quark lines.
}
\label{f:blobbles}
\end{figure}

To express the required momentum derivatives of the three-point correlation function, 
we identify the point-split vector current vertex
\begin{equation}
\frac{\partial D (\vec{\theta} \, )}{\partial p_i} \Big|_{\vec{\theta} = \vec{0}}
=
i V_i
,\end{equation}
\begin{widetext}
\noindent
which has the coordinate-space matrix elements
\begin{eqnarray}
\langle x' | V_\mu | x \rangle
&=&
\frac{1}{2}
\Big[
\delta_{x'+\mu, x}
(\gamma_\mu  - 1)
\cU_\mu (x')
+
\delta_{x',x+\mu}
(\gamma_\mu + 1)
\cU^\dagger_\mu (x)
\Big]
.\end{eqnarray}
The mean-square charge radius can be extracted by computing the ratio in
Eq.~\eqref{eq:r2},
with
\begin{eqnarray}
\frac{
\partial^2 \cC_4 ( \vec{p} \, ', \vec{p} \, )
}
{\partial p'_1 \partial p_1} \Big|_{\vec{p}\,' = \vec{p} = \vec{0}
}
&=&
-
\sum_{\vec{x}, \vec{y}}
\Bigg\langle
\Tr
\Big[
\gamma_5 
\langle x | S V_1 S | y \rangle
V_4 (y)
\langle y | S V_1 S | 0 \rangle
\gamma_5 
\langle 0 | S | x \rangle 
\Big]
\Bigg\rangle
\label{eq:D1D1}
,\end{eqnarray}
by virtue of the modified three-point function appearing in 
Eq.~\eqref{eq:master}. 
In the above expression, 
we have replaced the temporal component of the current with its point-split form. 
The correlator in 
Eq.~\eqref{eq:D1D1}
is computed with vanishing initial- and final-state momenta, 
and is graphically depicted in  
Fig.~\ref{f:blobbles}. 
For the 
$<r^4>$
moment of the charge distribution, 
one would need to compute
\begin{eqnarray}
\frac{
\partial^4 \cC_4 ( \vec{p} \, ', \vec{p} \, )}{\partial p'_1 \partial p_1 \partial p'_2 \partial p_2} 
\Big|_{\vec{p}\,' = \vec{p} = \vec{0}}
=
2
\Big\langle
\Tr
\Big[
\gamma_5 
S V_1 
S V_2 
S V_4 
S V_1 
S V_2
S 
\gamma_5
S
\Big]
+
\Tr
\Big[
\gamma_5 
S V_1 
S V_2 
S V_4 
S V_2 
S V_1
S 
\gamma_5
S
\Big]
\Big\rangle
\label{eq:DDDDC}
,\end{eqnarray}
and determine the ratio in Eq.~\eqref{eq:r4}. 
We have now dropped all coordinate dependence for ease, 
and have appealed to cubic symmetry to combine terms. 
Finally the 
$<r^6>$
moment requires evaluating the correlation functions
\begin{eqnarray}
\frac{
\partial^6 \cC_4 ( \vec{p} \, ', \vec{p} \, )
}
{
\partial p'_1 \partial p_1 \partial p'_2 \partial p_2 \partial p'_3 \partial p_3
} 
\Big|_{\vec{p}\,' = \vec{p} = \vec{0}}
&=&
-6
\sum_{\cP(1,2,3)}
\Big\langle
\Tr
\Big[
\gamma_5 
S V_1 
S V_2 
S V_3
S V_4 
S V_{\cP_1} 
S V_{\cP_2}
S V_{\cP_3}
S 
\gamma_5
S
\Big]
\Big\rangle
\label{eq:DDDDDDC}
,\end{eqnarray}
\end{widetext}
where the sum is over all permutations 
$\cP$
of 
$1$, $2$, $3$, 
which we write as
$\cP(1,2,3) = ( \cP_1, \cP_2, \cP_3)$. 
Cubic symmetry has been employed to reduce the number of terms appearing in the above expression.

%
%
%
%
%
%

A  cause for concern about the zero momentum method arises from problems encountered at finite lattice spacing.%
\footnote{
We thank H.~Wittig for alerting us to this possibility.
} 
In the context of the hadronic vacuum polarization, 
for example, 
twisted boundary conditions are known to introduce contact terms that diverge with an inverse power of the lattice spacing in the continuum limit%
~\cite{Aubin:2013daa,Horch:2013lla}.
Such contact terms exist in the product of two vector currents corresponding to the same direction,
whereas the divergent contribution is absent for orthogonal directions. 
The above expressions for various derivatives of the temporal current matrix element, 
Eqs.~\eqref{eq:D1D1}--\eqref{eq:DDDDDDC},  
are written in terms of multiple insertions of the vector current and therefore appear subject to the same complication. 
Our use of twisted boundary conditions, 
however, 
is only a theoretical device to obtain these Taylor coefficients, 
which are then evaluated at vanishing twist. 
We must still worry that differentiation could produce twist-independent terms that diverge in the continuum limit, 
however, 
we can show that the momentum derivatives required in this particular application prohibit such contributions.

The starting point for our application of the zero momentum method is the temporal current matrix element computed with twisted 
initial- and final-state quarks. 
Because the twists are taken in the spatial directions, 
there can be no additional divergent contribution to the 
\emph{temporal} 
current matrix element itself. 
Upon differentiation of this matrix element, 
divergent contact terms will be introduced. 
For example, 
the two derivatives with respect to 
$x$-components of momentum 
required to obtain the charge radius in
Eq.~\eqref{eq:D1D1}, 
will produce a divergent contribution analogous to that in the hadronic vacuum polarization. 
Evaluation at vanishing $x$-component of twists, 
however, 
removes this divergent contribution. 
To arrive at 
Eq.~\eqref{eq:DDDDC}, 
we need further differentiation, 
but with respect to the 
$y$-components of momentum. 
This produces another divergent contribution, 
which vanishes when the $y$-components of twists are set to zero. 
Finally differentiation with respect to the remaining 
$z$-components of momentum is necessary to evaluate 
Eq.~\eqref{eq:DDDDDDC},
and produces a power-divergent term which vanishes when the 
$z$-components of twists are set to zero.

There is a stronger argument against such power-law divergences
that holds even in the absence of twisted boundary conditions. 
In the evaluation of any of the three modified correlation functions, 
identical currents are never at the same temporal location. 
This owes directly to the fact that we utilize the temporal current matrix element, 
and differentiate with respect to the spatial momenta carried by intitial- and final-state quarks, 
which produces spatial vector-current vertices. 
In the absence of contact terms, 
three moments of the pion's charge distribution can thus be accessed without the need to perform subtractions of divergent contributions. 
If one employs rest-frame or Breit-frame matrix elements, 
however,
we expect power-law divergences will be encountered in the continuum limit. 
This complication arises in addition to the power-law Euclidean time dependence introduced upon momentum differentiation of the matrix element in 
Eq.~\eqref{eq:VCCF}.

\section{Finite Volume Corrections}      \label{s:FVC}                                %

With Eqs.~\eqref{eq:r2}--\eqref{eq:r6}, 
one can determine moments of the pion's charge distribution at vanishing initial-
and 
final-state momenta. 
As the method has been devised to circumvent the restriction to periodic momentum modes on a lattice of fixed size, 
it is susceptible to finite volume effects that are potentially substantial. 
We use a modification of chiral perturbation theory to ascertain the size of finite volume corrections 
on the required momentum derivatives of three-point functions.

\subsection{Partially Twisted Chiral Perturbation Theory}    %

To compute the required momentum derivatives of three-point correlation functions on a lattice of fixed size, 
we require partially twisted boundary conditions. 
In considering derivatives of the three-point function, 
we are forced to distinguish between an initial-state quark, 
a final-state quark,
and a spectator quark. 
These quarks, 
moreover, 
are all valence quarks 
because the current self-contraction does not contribute to pion matrix elements. 
Based on these observations, 
we consider the partially quenched QCD Lagrange density
\begin{equation}
\cL
= 
\ol Q ( \Dslash + m_Q ) Q
\label{eq:PQQCD}
,\end{equation}
for the graded flavor group
$SU(5|3)$.
The fundamental vector 
$Q$
accommodates eight flavors of quark fields
\begin{equation}
Q_i
=
\begin{pmatrix} 
u_1, &
u_2, &
d, &
j, &
l, &
\tilde{u}_1, &
\tilde{u}_2, &
\tilde{d} \,
\end{pmatrix}_i
.\end{equation} 
We employ the isospin limit in the valence and sea sectors.
Accordingly the quark mass matrix has the form
\begin{equation}
m_Q 
= 
\diag 
\left( 
m_v, 
m_v, 
m_v, 
m_{\text{sea}}, 
m_{\text{sea}}, 
m_v, 
m_v, 
m_v
\right)
,\end{equation} 
with 
$m_v$
the valence quark mass (which must be degenerate with the ghost quark mass), 
and 
$m_{\text{sea}}$
the sea quark mass. 
Even at the unitary mass point, 
$m_{\text{sea}} = m_v$, 
the theory remains partially quenched by virtue of differing boundary conditions on the various quark fields. 
This we refer to as partially twisted.

To handle the quark boundary conditions, 
we take the 
$Q$
field to be periodic, 
with the effect of partial twisting relegated to the gauge-covariant derivative. 
This derivative includes a term from uniform gauge fields
\begin{equation}
D_\mu = D_\mu^{SU(3)} + i B_\mu
,\end{equation} 
where 
$D_\mu^{SU(3)}$
is the QCD gauge-covariant derivative, 
and 
$B_\mu$
encodes the uniform 
$U(1)$
gauge fields.  
The uniform fields have vanishing temporal component, 
$B_\mu = ( \vec{B}, 0 )$, 
with the spatial components specified by
\begin{equation}
B_i
= 
\frac{1}{L}
\diag
\left( 
\theta_i, 
\theta'_i, 
0, 
0, 
0, 
\theta_i, 
\theta_i', 
0
\right)
.\end{equation}
From this choice, 
we see that the quark 
$u_1$
corresponds to the initial-state quark, 
while 
$u_2$
corresponds to the final-state quark. 
The down quark, 
$d$, 
is the spectator quark. 
The active and spectator quarks are valence quarks. 
For this reason, 
the theory is graded: 
we include ghost quarks
$\tilde{u}_1$, 
$\tilde{u}_2$, 
and
$\tilde{d}$
in order for this identification to be made. 
Finally the quarks 
$j$
and 
$l$
are the sea quarks, 
which are not twisted. 
To induce momentum transfer, 
the vector current is taken to correspond to a flavor-changing current 
\begin{equation}
V_\mu = \ol u_2 \gamma_\mu u_1
\label{eq:V}
,\end{equation}
within the extended flavor group.

To address finite volume effects, 
we consider the low-energy theory of pions. 
The low-energy effective theory of partially quenched QCD
is partially quenched chiral perturbation theory%
~\cite{Bernard:1993sv,Sharpe:2000bc,Sharpe:2001fh}. 
This theory describes the infrared dynamics of the low-energy pseudoscalar modes emerging from 
spontaneous breaking of chiral symmetry. 
Our considerations are that of the 
$p$-regime of finite volume chiral perturbation theory%
~\cite{Gasser:1987zq}, 
in which the zero modes of the Goldstone manifold remain weakly coupled. 
Chiral perturbation theory is written in terms of the coset field,
$\Sigma = \exp ( 2 i \Phi / f )$, 
where 
$\Phi$
is a matrix of pseudoscalar meson fields. 
The dimensionful parameter 
$f$
can be identified as the pion decay constant, 
which has the numerical value
$f \approx 132 \, \texttt{MeV}$
in our conventions.

To obtain partially quenched chiral perturbation theory corresponding to the Lagrange density in 
Eq.~\eqref{eq:PQQCD}, 
we use the coset transformation law
$\Sigma \to L \Sigma R^\dagger$
under a graded chiral transformation, 
$(L, R) \in U(5|3)_L \otimes U(5|3)_R$. 
The effective theory is then written down by enumerating all chirally invariant operators. 
Sources of explicit chiral symmetry breaking are additionally included using the spurion trick. 
For our application to the pion form factor with partially twisted boundary conditions, 
we must incorporate the vector current
and the uniform gauge fields, 
in addition to the quark mass. 
In terms of the coset field 
$\Sigma$, 
the leading-order Lagrange density has the form
\begin{equation}
\cL
= 
\frac{f^2}{8}
\str \left( D_\mu \Sigma^\dagger D_\mu \Sigma \right)
- 
\lambda \str \left( m_Q \Sigma^\dagger +  m_Q  \Sigma \right)
+ 
\mu_0^2 \Phi_0^2
.\end{equation}
The field 
$\Phi_0$
is the flavor-singlet field, 
$\Phi_0 = \str ( \Sigma) / \sqrt{2}$, 
and has been included only as a device to obtain the flavor-neutral propagators. 
The covariant derivative
$D_\mu$
is specified by its action on the coset field
\begin{equation}
D_\mu \Sigma
= 
\partial_\mu \Sigma
+ 
i A_\mu [ T, \Sigma]
+ 
i [ B_\mu, \Sigma] 
,\end{equation}
where 
$A_\mu$
is an external vector field, 
and the matrix 
$T$
is given by
\begin{equation}
T_{ij} 
= 
\delta_{i2} \delta_{j1}
,\end{equation}
which corresponds to the flavor structure of the flavor-changing vector current in 
Eq.~\eqref{eq:V}.

The meson fields are embedded in 
$\Sigma$
through 
$\Phi$
which takes the form of an
eight-by-eight matrix, 
which can be written in blocks
\begin{equation}
\Phi
=
\begin{pmatrix}
M_{vv} & M_{vs} & \chi^\dagger_{gv} \\
M_{sv} & M_{ss} & \chi^\dagger_{gs} \\
\chi_{gv} & \chi_{gs} & M_{gg}
\end{pmatrix}
.\end{equation}
The blocks
$M_{vv}$
and 
$M_{gg}$
contain bosonic mesons composed of a quark-antiquark pair, 
and a ghost quark-ghost antiquark pair, 
respectively. 
Such mesons appear in these blocks as
\begin{equation}
M_{vv} 
= 
\begin{pmatrix}
\eta_{11} & \eta_{12} & \pi^+_1 \\
\eta_{21} & \eta_{22} & \pi^+_2 \\
\pi^-_1 & \pi^-_2 & \eta_d 
\end{pmatrix}
,\end{equation}
and
\begin{equation}
M_{gg} 
= 
\begin{pmatrix}
\tilde{\eta}_{11} & \tilde{\eta}_{12} & \tilde{\pi}^+_1 \\
\tilde{\eta}_{21} & \tilde{\eta}_{22} & \tilde{\pi}^+_2 \\
\tilde{\pi}^-_1 & \tilde{\pi}^-_2 & \tilde{\eta}_d 
\end{pmatrix}
.\end{equation}
The 
$\eta_{ij}$ 
mesons have quark content
$\eta_{ij} \sim u_i  \ol{u}_j$, 
whereas the 
$\tilde{\eta}_{ij}$ 
mesons have quark content
$\tilde{\eta}_{ij} \sim \tilde{u}_i \overline{\tilde{u}}_j$. 
The neutral meson composed of a spectator quark-antiquark pair is 
$\eta_d \sim d \ol d$, 
while the ghostly counterpart is 
$\tilde{\eta}_d \sim \tilde{d} \, \overline{ \tilde{d}}$. 
The electrically charged mesons have been denoted with 
$\pi$'s,
specifically
$\pi^+_i \sim  u_i \ol d$
and
$\tilde{\pi}^+_i \sim \tilde{u}_i \overline{\tilde{d}}$. 
The valence-sea and sea-sea mesons are also bosonic. 
These degrees of freedom appear in the matrices
$M_{vs}$
and
$M_{ss}$, 
with
\begin{equation}
M_{vs} 
= 
\begin{pmatrix}
\phi_{j u_1} & \phi_{j u_2} & \phi_{j d} \\
\phi_{l u_1} & \phi_{l u_2} & \phi_{l d} \\
\end{pmatrix}
,\end{equation}
and
\begin{equation}
M_{ss} 
= 
\begin{pmatrix}
\eta_j & \pi_{jl} \\
\pi_{lj} & \eta_l 
\end{pmatrix}
,\end{equation}
respectively. 
The remaining mesons are fermionic fields. 
Those composed of a ghost quark and a valence antiquark are contained in the matrix 
$\chi_{gv}$, 
while those composed of a ghost quark and a sea antiquark appear in 
$\chi_{sv}$. 
These matrices of fermionic mesons have the form
\begin{equation}
\chi_{gv} 
= 
\begin{pmatrix}
\phi_{\tilde{u}_1 u_1} & \phi_{\tilde{u}_1 u_2} & \phi_{\tilde{u}_1 d} \\
\phi_{\tilde{u}_2 u_1} & \phi_{\tilde{u}_2 u_2} & \phi_{\tilde{u}_2 d} \\
\phi_{\tilde{d} u_1} & \phi_{\tilde{d} u_2} & \phi_{\tilde{d} d} 
\end{pmatrix}
,\end{equation}
and
\begin{equation}
\chi_{sg} 
= 
\begin{pmatrix}
\phi_{\tilde{u}_1 j} & \phi_{\tilde{u}_1 l} \\
\phi_{\tilde{u}_2 j} & \phi_{\tilde{u}_2 l} \\
\phi_{\tilde{d} j} & \phi_{\tilde{d} l} 
\end{pmatrix}
.\end{equation}

Treating the Lagrange density to tree level, 
we see that the quark-basis mesons with quark content 
$Q$ 
and 
antiquark content
$Q'$
have masses given by
\begin{equation}
m^2_{QQ'} 
= 
\frac{\lambda}{4f^2}
(m_Q + m_{Q'})
.\end{equation}
The flavor-singlet meson, 
$\Phi_0$,
has an additional contribution to its mass proportional to 
$\mu_0^2$. 
Because the flavor-singlet axial symmetry of the partially quenched theory is anomalous, 
we can integrate the flavor singlet out of partially quenched chiral perturbation theory%
~\cite{Sharpe:2001fh}. 
The resulting Goldstone manifold becomes 
$SU(5|3)_L \otimes SU(5|3)_R / SU(5|3)_V$, 
however, 
the flavor-neutral mesons, 
which appear on the diagonal of the matrix
$\Phi$,
do not have simple Klein-Gordon propagators. 
Instead, 
their propagators have both quark-connected and quark-disconnected terms, 
where the latter are conventionally referred to as hairpins. 
The propagator matrix between 
$\eta_a$-$\eta_b$
quark-basis states%
\footnote{
Due to our notation for the various mesons, 
one must be careful to note that this propagator matrix only applies to the diagonal entries of the matrix 
$\Phi$. 
These states are the flavor-neutral mesons, 
whereas other 
$\eta$ states that are off the diagonal, 
such as the 
$\eta_{12}$,
are only electrically neutral rather than flavor neutral. 
} 
is given by
\begin{equation}
G_{\eta_a \eta_b} (k)
= 
\frac{\epsilon_a \delta_{ab}}{k^2 + m_{\eta_a}^2}
- 
\frac{1}{2}
\frac{\epsilon_a \epsilon_b (k^2+m_{\eta_j}^2)}{(k^2 + m_{\eta_a}^2)(k^2 + m_{\eta_b}^2)}
.\end{equation}
The first thing to note is that twisting does not affect the propagation of flavor-neutral states. 
For this reason, 
the twist angles
$\vec{\theta}$
and
$\vec{\theta} \,'$
are absent from the flavor-neutral propagator matrix. 
In partially quenched chiral perturbation theory, 
the lack of unitarity of the theory arises from the disconnected part of the flavor-singlet propagator; 
however, 
the double pole is absent with degenerate valence and sea quark masses.%
\footnote{
In generalizing to include the strange quark, 
the result remains true. 
Because flavor-neutral states are unaffected by twisting, 
the double-pole sickness also goes away when valence quarks are made degenerate with their sea quark counterparts. 
In this limit, 
the lack of unitarity of the partially twisted theory does not lead to any enhancement of the finite volume effects.
}

The propagators of flavor non-singlet mesons have a simple Klein-Gordon form, 
however, 
they acquire dependence on the twist angles. 
Writing the scalar propagator in the form 
\begin{equation}
G_m(\vec{k}, k_4)
= 
\frac{1}{\vec{k} \, {}^2 + k_4^2  + m^2}
,\end{equation}
we see that a meson with quark content 
$Q$
and antiquark content
$Q'$, 
with 
$Q' \neq Q$,
is described by the propagator 
\begin{equation}
G_{m_{QQ'}} 
(\vec{k} + \vec{B}_Q - \vec{B}_{Q'}, k_4)
,\end{equation}
where 
$\vec{k} = \frac{2 \pi \vec{n}}{L}$
is a periodic momentum mode.

\subsection{Pion Current Matrix Element in an Arbitrary Frame}    %

%
%
%
\begin{figure}
\includegraphics[width=0.3\textwidth]{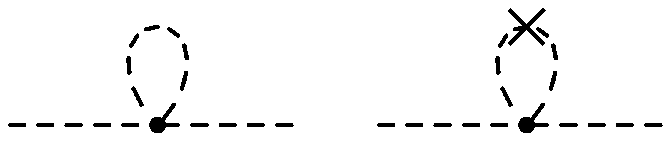}%
\\
\bigskip
\bigskip
\bigskip
\includegraphics[width=0.3\textwidth]{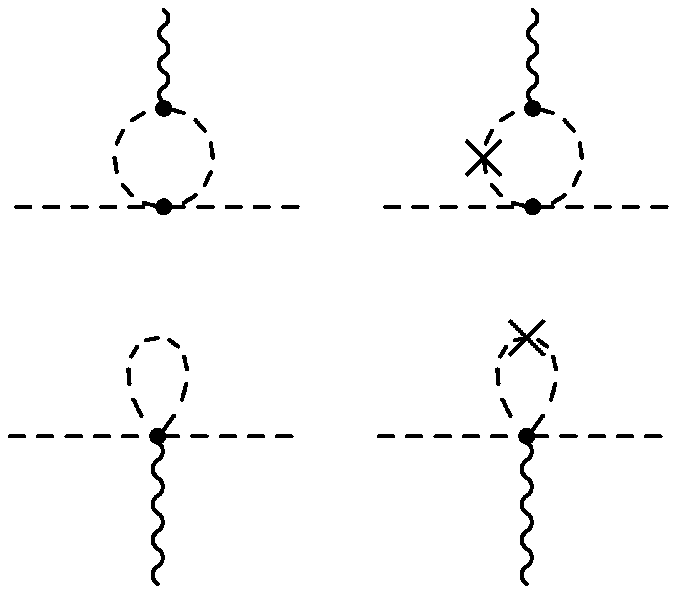}%
\caption{
One-loop contributions to the pion current matrix element in partially quenched chiral perturbation theory. 
The top panel shows diagrams required to determine the pion wavefunction renormalization. 
Dashed lines represent mesons, 
while the cross represents the partially quenched hairpin interaction.  
The wiggly line shows the insertion of the vector current, 
which corresponds to a flavor-changing transition in the effective theory. 
The flavors differ only by their boundary conditions. 
}
\label{f:1loop}
\end{figure}
%
%
%

Having spelled out the necessary ingredients of partially quenched chiral perturbation theory, 
we now detail the computation of the pion electromagnetic form factor. 
The form factor can be determined in the effective theory by computing the matrix element
\begin{eqnarray}
\cM_\mu ( \vec{p} \, ', \vec{p} \, )
&\equiv&
\langle \pi_2^+ (\vec{0} \,) | V_\mu | \pi_1^+ ( \vec{0} \, ) \rangle_L
\notag \\
&\overset{L \to \infty}{\longrightarrow}&
\langle \pi^+ (\vec{p} \, ') | V_\mu | \pi^+ ( \vec{p} \, ) \rangle
\label{eq:VMEL}
.\end{eqnarray}
On the first line, 
we label the matrix element of pions by their vanishing Fourier momentum. 
On the second line, 
we show the infinite volume limit of the matrix element. 
By virtue of partial twisting, 
the initial-state pion carries momentum 
$\vec{p} = \frac{\vec{\theta}}{L}$,
and the final-state pion carries momentum 
$\vec{p} \, ' = \frac{\vec{\theta} \, '}{L}$.
In taking the infinite volume limit, 
it is assumed that
$\vec{p}$ 
and
$\vec{p} \, '$
are held fixed.  
These momenta can be continuously varied on a lattice of fixed size,
and enable us to perform the Taylor series expansion of current correlation functions in chiral perturbation theory. 
The momentum derivatives required to evaluate moments of the charge distribution at vanishing momentum,  
Eqs.~\eqref{eq:r2}--\eqref{eq:r6}, 
can all be derived from the pion current matrix element calculated in an arbitrary frame. 
For this reason, 
we determine the general expression for the finite volume effect in an arbitrary frame in this section, 
and evaluate the required Taylor series coefficients in the following section.%
\footnote{
In passing, 
we note an equivalent alternate method. 
One can compute the finite volume corrections by momentum expanding the meson propagators and vertices appearing in partially twisted chiral perturbation theory. 
The coefficients of the Taylor series expansion are then computed directly by using appropriately modified correlation functions at zero momentum in the effective theory. 
For the three moments of the charge distribution determined above, 
this would mean three separate, 
though related, 
computations. 
We find it easier to evaluate the original current matrix element at finite twist angles, 
and derive all three results from differentiation.  
}

To compute the finite volume corrections to the current matrix element in 
Eq.~\eqref{eq:VMEL}, 
we must evaluate the one-loop diagrams shown in 
Fig.~\ref{f:1loop}. 
Due to rather fortuitous cancellations, 
there are no hairpin contributions to the current matrix element. 
Additionally complete cancellation of valence-valence meson contributions against those of valence-ghost mesons leads to dependence only on the valence-sea meson mass, 
which is known to occur for the infinite volume partially quenched pion form factor%
~\cite{Arndt:2003ww}.  
For the sake of notational ease, 
we denote the mass of valence-sea mesons simply by 
$m$. 
The finite volume effect on the temporal component of the current matrix element
is defined by
\begin{equation}
\Delta \cM_4
= 
\cM_4(L)
- 
\cM_4(L = \infty)
.\end{equation}
Evaluation of the diagrams shown in the figure can be cast in the general form
\begin{eqnarray} \label{eq:FFFV}
\Delta \cM_4
&=&
(p'_4 + p_4)
\, \Delta F
+ 
q_4 
\, \Delta G
,\end{eqnarray}
where 
$\Delta F$
and
$\Delta G$
represent frame-dependent finite volume corrections to the current matrix element. 
The latter contribution, 
$\Delta G$,  
is forbidden in infinite volume due to current conservation;
however, 
in a finite volume with twisted boundary conditions, 
this contribution becomes essential to maintain the Ward-Takahashi identity, 
see%
~\cite{Bijnens:2014yya}.

While the explicit form of 
$\Delta G$
is rather lengthy in an arbitrary frame, 
it will fortunately not be needed for our computation.  
To see this, 
we notice that the momentum derivatives of the current matrix element which are required to determine the charge radius at zero momentum have the form
\begin{equation} \label{eq:ddM}
\frac{\partial^2}{\partial p'_1 \partial p_1}
\Delta \cM_4
\Big|_{\vec{p} \, ' = \vec{p} = \vec{0}}
,\end{equation}
with further momentum derivatives of the matrix element needed to obtain the two higher moments of the charge distribution. 
The crucial observation is that the 
$\Delta G$
term of the current matrix element in 
Eq.~\eqref{eq:FFFV}
drops out of these momentum derivatives when evaluated at zero initial- and final-state momenta. 
Focusing on the differentiation required for the charge radius, 
we have
\begin{equation} \label{eq:ddG}
\frac{\partial^2}{\partial p'_1 \partial p_1}
q_4 \, \Delta G 
=
\frac{\partial p'_4}{\partial p'_1}
\frac{\partial \Delta G}{\partial p_1} 
-
\frac{\partial p_4}{\partial p_1}
\frac{\partial \Delta G}{\partial p'_1} 
+
q_4
\frac{\partial^2 \Delta G }{\partial p'_1 \partial p_1}
.\end{equation}
Because the current matrix element must be well behaved at 
$\vec{p} = \vec{0}$
and at 
$\vec{p} \, ' = \vec{0}$, 
the first term on the right-hand side above vanishes when evaluated at 
$p'_1 = 0$, 
the second term vanishes when evaluated at 
$p_1 = 0$, 
and the last term vanishes when evaluated at
$\vec{p} \, ' = \vec{p}$. 
Consequently the finite volume effect proportional to
$q_4 \, \Delta G$
will not contribute to the derivatives required to determine the charge radius at zero momentum, 
$\vec{p} \, ' = \vec{p} = \vec{0}$. 
The same conclusion can easily be reached for the two higher moments of the charge distribution. 
One merely takes further momentum derivatives of Eq.~\eqref{eq:ddG}, 
and observes that all terms produced necessarily vanish when evaluated at
vanishing initial- and final-state momenta.

To compute the effect of finite volume on the method of extracting moments of the charge distribution at zero momentum, 
we thus need only the function 
$\D F$
appearing in the finite volume current matrix element, 
Eq.~\eqref{eq:FFFV}. 
Taking contributions from all one-loop diagrams, 
we find
\begin{eqnarray}
\Delta F
&=&
\frac{1}{f^2}
\int_0^1 dx \, 
\mathcal{I}_{\frac{1}{2}}
\left[
x \vec{p} \, ' + (1-x) \vec{p}, 
m^2
+
x(1-x) q^2
\right]
\notag \\
&&
-
\frac{1}{2 f^2}
\left[
\mathcal{I}_{\frac{1}{2}} (\vec{p}\, ', m)
+ 
\mathcal{I}_{\frac{1}{2}} ( \vec{p}, m)
\right]
\label{eq:DF}
,\end{eqnarray}
This finite volume effect has been compactly written in terms of the basic functions
\begin{eqnarray}
\mathcal{I}_\b
(\vec{p}, m)
&\equiv&
\frac{1}{L^3}
\sum_{\vec{n}}
\frac{1}{\left[\left( \frac{2 \pi}{L} \vec{n} + \vec{p} \, \right)^2+m^2\right]^\b}
\notag \\
&&-
\int \frac{d \vec{k}}{(2\pi)^3}
\frac{1}{\left[ \left(\vec{k} + \vec{p} \, \right)^2  + m^2\right]^\b}
\label{eq:I}
,\end{eqnarray}
which encompass the difference between the finite volume mode sum, 
and in the infinite volume momentum integration. 
In taking derivatives with respect to 
$p_i$, 
removing the 
$\vec{p}\,$
from the integral by translating the momentum integration over
$\vec{k}$
leads to spurious infinite volume contributions. 
This peculiarity is detailed in the appendix. 
The functions defined above can be recast in terms of Jacobi elliptic-theta functions%
~\cite{Sachrajda:2004mi,Jiang:2008ja}, 
for which we have
\begin{eqnarray}
\mathcal{I}_\beta (\vec{p}, m)
&=&
\frac{1}{(4 \pi)^{\frac{3}{2}} \Gamma(\beta)}
\int_0^\infty ds \, s^{ \b - \frac{5}{2}} e^{ - s m^2}
\notag \\
&& 
\times
\left[
\prod_{j=1}^3
\vartheta_3 \left(\frac{L p_j}{2} ,e^{ - \frac{L^2}{4 s}} \right)
-
1
\right]
\label{eq:Poissond}
.\end{eqnarray}
In working at zero momentum, 
we will often employ the abbreviated notation 
$\mathcal{I}_\b (m) \equiv \mathcal{I}_\b ( \vec{0}, m)$.

The expression for the finite volume effect
$\D F$
in Eq.~\eqref{eq:DF}
can be checked against results known in various limits. 
First notice that in the forward limit, 
we have
\begin{equation}
\D F \Big|_{\vec{p} \, ' = \vec{p} } = 0
,\end{equation} 
which consequently leads to a vanishing volume correction to the time component of the current,
$\D \cM_4 = 0$. 
This corresponds to non-renormalization of the pion charge due to finite volume,
and is a consequence of the Ward identity. 
The Ward identity happens to be satisfied for the time component of the current due to our specification of an infinite 
temporal extent, 
see%
~\cite{Hu:2007eb}. 
Evaluating the finite volume effect
$\Delta F$ 
in the pion rest frame, 
which is specified by setting
$\vec{p} = \vec{0}$
with 
$\vec{q} = \vec{p} \, '$, 
we recover the corresponding result for 
$\Delta F$
determined from isospin twisted boundary conditions%
~\cite{Jiang:2006gna}. 
To compare with the latter result, 
we must set the down quark twist to zero in order for the setup to match that used in the present work. 
Finally
we can use our result for 
$\Delta F$
in an arbitrary frame to determine the finite volume effect in the Breit frame, 
which is specified by the kinematics
$\vec{p} = - \frac{1}{2} \vec{q}$, 
and 
$\vec{p} \, ' = \frac{1}{2} \vec{q}$. 
With this choice, 
one obtains from 
$\Delta F$
in 
Eq.~\eqref{eq:DF}
the result of%
~\cite{Jiang:2008te}.

\subsection{Finite Volume Corrections at Zero Momentum}     %

To evaluate the effect of finite volume on determining form factors at zero momentum, 
we return to the vector-current correlation function in Eq.~\eqref{eq:VCCF}. 
We detailed the expected behavior of this correlation function in infinite volume, 
and now we consider the modifications arising in finite volume. 
Generalizing this correlation function to finite volume, 
we have
\begin{align}
\mathfrak{C}_4
( \vec{p} \, ', \vec{p}  \, | x_4, y_4)
=
|\mathcal{Z}|^2
\cM_4(\vec{p} \, ', \vec{p} \,)
\frac{
e^{ - \cE\left(\vec{p} \, ' \right) (x_4 - y_4)}
e^{- \cE\left(\vec{p} \, \right) y_4}
}
{
2 \cE\left( \vec{p} \, '\right) 
2 \cE\left( \vec{p} \, \right)
}
.\notag \\
\label{eq:FVCC}
\end{align}
The overlap factor in finite volume is denoted by 
$\mathcal{Z}$, 
which is momentum independent. 
Additionally appearing in the above expression is the temporal component of the finite volume current matrix element, 
$\cM_4(\vec{p} \, ', \vec{p} \,)$, 
as well as the charged pion energy in finite volume,
$\cE (\vec{p} \, )$, 
which has been determined using partially twisted chiral perturbation theory in%
~\cite{Jiang:2006gna}. 
For our application, 
it is important to note that
$\cE (\vec{p} \,)$
has a regular expansion about vanishing twist,
$\vec{p} = \vec{0}$, 
i.e.~it can be written in the form
\begin{equation} \label{eq:FM}
\cE(\vec{p} \, )^2 
=
\vec{p} \,{}^2 + m^2 + \Delta m^2 (\vec{p} \,)
,\end{equation}
with 
$\Delta m^2 (\vec{p} \,) = \delta m^2 + \vec{p} \, {}^2  \delta p^2+ \cdots$.
Here, 
$\Delta m^2 (\vec{p} \,)$
is the finite volume correction to the mass. 
It retains twist-angle dependence, 
and so can either be thought of as a momentum-dependent finite volume mass shift%
~\cite{Bijnens:2014yya}, 
or as arising from renormalization of the pion momentum%
~\cite{Jiang:2006gna}.
From the twist-angle expansion in Eq.~\eqref{eq:FM}, 
we see
$\partial \cE (\vec{p} \, ) / \partial p_j \big|_{p_j = 0}= 0$, 
and can conclude that the required momentum derivatives acting on the finite volume correlation function, 
Eq.~\eqref{eq:FVCC},  
all act on 
$\cM_4 ( \vec{p} \, ', \vec{p} \, )$,
or else evaluate to zero at vanishing momentum.

In light of these observations, 
we can ascertain the finite volume corrections to the zero-momentum method. 
First, 
we define the finite volume effect on the charge radius, 
$\Delta r^2$,  
using the finite volume current correlation functions
\begin{equation}
\frac{3}{\mathfrak{C}_4 ( \vec{0}, \vec{0})} 
\frac{\partial^2 \mathfrak{C}_4 (\vec{p} \, ', \vec{p} \, )}{\partial p'_1 \partial p_1}
\Big|_{\vec{p} \, ' = \vec{p} = \vec{0}}
=
< r^2 > 
+ 
\Delta r^2 
.\end{equation}
Recalling the form of 
Eqs.~\eqref{eq:r4} and \eqref{eq:r6}, 
the finite volume effect on the two higher moments, 
$\Delta r^4$
and
$\Delta r^6$,
can then be defined in a completely analogous fashion. 
Taking the required derivatives of the finite volume current correlation function in Eq.~\eqref{eq:FVCC}, 
and using the finite volume current matrix element in Eq.~\eqref{eq:FFFV}, 
we arrive at the simple expressions
\begin{eqnarray}
\Delta r^2 
&=&
3
\frac{\partial^2 \Delta F}{\partial p'_1 \partial p_1}
\Big|_{\vec{p} \, ' = \vec{p} = \vec{0}},
\\
\Delta r^4
&=&
45
\frac{\partial^4 \Delta F}{\partial p'_1 \partial p_1 \partial p'_2 \partial p_2}
\Big|_{\vec{p} \, ' = \vec{p} = \vec{0}},
\\
\Delta r^6 
&=&
1575
\frac{\partial^6 \Delta F}{\partial p'_1 \partial p_1 \partial p'_2 \partial p_2 \partial p'_3 \partial p_3}
\Big|_{\vec{p} \, ' = \vec{p} = \vec{0}}
.\end{eqnarray}
Finally, 
we can evaluate these finite volume corrections by explicitly taking derivatives of the result determined from 
chiral perturbation theory,  
Eq.~\eqref{eq:DF}.

To expedite computation of the derivatives, 
we notice the following property of the energy denominator
appearing in the finite volume mode sum and momentum integral
\begin{eqnarray}
[\vec{k} + x \vec{p} \, ' + ( 1 - x) \vec{p} \, ]^2
+ 
m^2 
+ 
x ( 1- x) q^2
= 
\notag  \\  
\vec{k}^2
+ 
2 \vec{k} \cdot [ x \vec{p} \, ' + ( 1- x) \vec{p} \, ]
+ 
m^2
+ 
f( \vec{p} \, '{}^2, \vec{p} \, {}^2)
,\end{eqnarray}
where 
$f(\vec{0}, \vec{0} ) = 0$. 
Because of the vanishing of 
$\partial f( \vec{p} \, '{}^2, \vec{p} \, {}^2) / \partial p_j \big|_{p_j = 0}$, 
and similarly for the derivative with respect to the final-state momentum,
$p'_j$,  
we can effectively replace 
$f ( \vec{p} \, '{}^2, \vec{p} \, {}^2) \longrightarrow 0$
when taking momentum derivatives and evaluating at zero momentum. 
The finite volume correction to the charge radius determined at zero momentum is 
given by
\begin{equation}
\D r^2
=
\frac{1}{2 f^2} 
\left[
\mathcal{I}_{\frac{3}{2}} (m)
-
m^2 \mathcal{I}_{\frac{5}{2}} (m)
\right]
,\end{equation}
while the finite volume corrections to the two higher moments take the form
\begin{eqnarray}
\D r^4
&=&
\frac{21}{2 f^2} 
\left[
\mathcal{I}_{\frac{5}{2}} (m)
-
2 m^2 \mathcal{I}_{\frac{7}{2}} (m)
+
m^4 \mathcal{I}_{\frac{9}{2}} (m)
\right],
\notag \\
\\
\D r^6
&=&
\frac{4455}{4 f^2} 
\Bigg[
\mathcal{I}_{\frac{7}{2}} (m)
-
3 m^2 \mathcal{I}_{\frac{9}{2}} (m)
\notag \\
&& \phantom{spac} +
3 m^4 \mathcal{I}_{\frac{11}{2}} (m)
-
m^6 \mathcal{I}_{\frac{13}{2}} (m)
\Bigg]
.\end{eqnarray}

\begin{figure}
\begin{center}
\epsfig{file=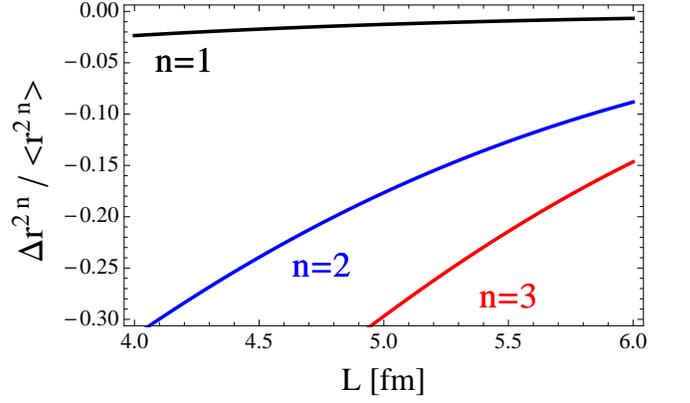,width=8.5cm}
\caption{
Finite volume effect on moments of the pion's charge distribution determined at zero momentum.
The relative differences 
$\Delta r^2 / < r^2>$, 
$\Delta r^4 / < r^4 >$, 
and 
$\Delta r^6 / < r^6>$
are plotted as a function of the lattice size
$L$.
The infinite volume value of the charge radius is taken from experiment, 
while values of the two higher moments are the one-loop predictions from chiral perturbation theory. 
In the plot, 
the pion mass is fixed to its physical value.}
\label{f:FV}
\end{center}
\end{figure}

To assess the finite volume effect on the radius, 
we will compare to the physical pion charge radius, 
$<r^2_\pi> = ( 0.67 \, \texttt{fm})^2$%
~\cite{Beringer:1900zz}. 
For the higher moments, 
we will use the corresponding infinite volume result determined from chiral perturbation theory at one-loop order. 
Using the dimensionally regulated form of the infinite volume integral in 
Eq.~\eqref{eq:DF}, 
we can write the form factor as
\begin{eqnarray}
F(q^2)
&=& 
1 
+ 
\frac{( 4 \pi \mu^2)^\epsilon}{8 \pi^2 f^2}
\int_0^1 dx 
\int_0^\infty \frac{ds}{s^{2 - \epsilon}}
e^{ - s m^2} 
\notag \\
&& \phantom{space} 
\times 
\left[ 
e^{ - s x(1-x) q^2} 
- 
1
\right]
.\end{eqnarray}
Carrying out subtraction of the 
$\epsilon$-pole in the 
$\ol{\text{MS}}$ 
scheme, 
we can identify the charge radius
\begin{equation}
<r^2> 
= 
\frac{1}{8 \pi^2 f^2} 
\left[ 
- 
\log \frac{m^2}{\mu^2}  
+ 
c(\mu) 
\right]
,\end{equation}
where
$c(\mu)$
is the coefficient of a local operator from the next-to-leading order chiral Lagrangian. 
This result is in agreement with%
~\cite{Gasser:1985ux}. 
Carrying out the momentum expansion further, 
and using the definition of the higher moments of the charge distribution in 
Eq.~\eqref{eq:moments},
we arrive at  
\begin{equation}
< r^{2n} >
=
\frac{n! (n-2)! (2 n -1)!!}{8 \pi^2 f^2 (m^2)^{n-1}}
,\end{equation}
for all 
$n > 1$. 
Using this one-loop result from chiral perturbation theory  gives us the values
$<r^4> = ( 0.77 \, \texttt{fm})^4$,
and 
$<r^6> =  (1.5 \, \texttt{fm})^6$.

The finite volume effect on moments of the pion's charge distribution is shown in Fig.~\ref{f:FV}. 
For the first three non-trivial moments, 
$n = 1$--$3$,  
we plot the finite volume effect normalized by the corresponding infinite volume value, 
$\Delta r^{2n} / < r^{ 2n} >$,
as a function of the size of the lattice. 
A clear trend toward larger volume effects as a function of 
$n$ 
is shown. 
This lines up with physical intuition. 
In the conventional evaluation of the current matrix element, 
greater momentum resolution is required to determine the higher moments. 
Such resolution requires prohibitively large volumes. 
In the method at zero momentum, 
this difficulty is mirrored by the rather complicated form of the modified correlation functions that must be computed. 
Due to the momentum differentiation, 
furthermore, 
the finite volume effects are considerably enhanced over those for the current matrix element. 
For comparison, 
the asymptotic behavior of the finite volume correction to the current matrix element has the form
\begin{equation}
\Delta \cM_4
\sim 
L^{-3/2} \, e^{ - m_\pi L}
,\end{equation}
while the finite volume effects on charge distribution moments have the asymptotic behavior
\begin{eqnarray}
\Delta r^2 
&\sim& 
L^{1/2} \, e^{ - m_\pi L},
\notag
\\
\Delta r^4 
&\sim&
L^{5/2} \, e^{ - m_\pi L},
\notag 
\\
\Delta r^6 
&\sim&
L^{9/2} \, e^{ - m_\pi L}
.\end{eqnarray}
Despite such enhancement of finite volume effects, 
the overall scaling remains exponential. 
Fortunately 
it appears that the determination of the charge radius at zero momentum is considerably insensitive to the volume.

\section{Summary} \label{s:S}              %

Above we investigate the novel method proposed by%
~\cite{deDivitiis:2012vs}
to overcome large extrapolations to zero momentum. 
Our particular concern is with the extension of the method to the case of radii, 
and the assessment of corresponding finite volume effects. 
For moments of the pion's charge distribution, 
we find that the modified correlation functions given in 
Eqs.~\eqref{eq:D1D1}--\eqref{eq:DDDDDDC}
cleanly isolate the desired quantities at vanishing momentum. 
These correlation functions require taking momentum derivatives with respect to initial- and final-state quarks. 
To address finite volume effects, 
we must understand how to arrive at the modified correlation functions on a torus. 
The zero-momentum method can be derived on a finite lattice of fixed size by treating the active quarks as subject to twisted boundary conditions. 
Because the twist angles are continuous parameters, 
a Taylor series expansion can be performed about vanishing twist. 
The coefficients of the Taylor series expansion correspond to modified correlation functions, 
and these are to be determined using lattice QCD calculations, 
as proposed by%
~\cite{deDivitiis:2012vs}. 
Such calculations require only periodic quark fields and are evaluated at zero momentum. 
Finite volume corrections can correspondingly be derived by formulating chiral perturbation theory for 
twisted active quarks. 
In this work, 
an expression for the finite volume effect on the temporal component of the pion current matrix element in an arbitrary frame using an extended
$SU(5|3)$ 
flavor group is derived. 
Differentiation and evaluation of this expression at vanishing twist angles leads to the finite volume effect on the extraction of the charge radius 
(and two higher moments)
at zero momentum.  
Our expectation from chiral perturbation theory is that the volume corrections to determining the charge radius at zero momentum will be quite small.

Straightforward generalization of the method will allow one to investigate moments of the electric and magnetic form factors of the nucleon, 
for example. 
Because of the extra spectator quark, 
chiral perturbation theory computations for the nucleon will require the flavor group to be extended to 
$SU(6|4)$, 
as was pursued for Breit frame computations in%
~\cite{Jiang:2008ja}. 
Because the magnetic form factor arises from the spatial component of the electromagnetic current, 
one generally expects larger volume corrections to magnetic observables. 
As the magnetic form factor drops out of the current matrix element at vanishing momentum transfer, 
an additional momentum derivative will be required over the electric case. 
This extra differentiation will lead to further enhancement of the finite volume effects on magnetic quantities. 
It would be advantageous to assess these volume effects using chiral perturbation theory. 
In general, 
care must be taken in the zero momentum method to avoid power-divergent contributions in the continuum limit, 
however, 
the approach advocated here avoids these contributions, 
as well as removes power-law Euclidean time dependence from the Taylor coefficients.
Finally we remark that we have been unable to find a generalization of the method for the case of disconnected current insertions. 
In such cases, 
both quarks and gluons are momentum carriers, 
and this feature ultimately seems to require a different approach.

\begin{acknowledgments}
This work is supported in part by a joint City College of New York--RIKEN/Brookhaven Research Center fellowship, 
a grant from the Professional Staff Congress of the CUNY,
and by the U.S.~National Science Foundation, under Grant No.~PHY$12$-$05778$.
We are grateful to  
T.~Izubuchi
for discussions. 
\end{acknowledgments}

\appendix
\section{Derivatives of Mode Sums}
\label{s:A}

For completeness, 
we detail how to take momentum derivatives of mode sums without introducing spurious infinite volume contributions. 
Consider the basic finite volume function 
$\mathcal{I}_\b(\vec{p} \,, m)$
defined in 
Eq.~\eqref{eq:I}. 
Its Taylor series expansion has the form
\begin{equation}
\mathcal{I}_\b(\vec{p} \,, m)
= 
\mathcal{I}_\b(m)
+
\frac{1}{2} \vec{p} \,{}^2
\mathcal{I}^{(2)}_\beta (m)
+ 
\cdots
.\end{equation}
Using the expression for 
$\mathcal{I}_\b(\vec{p} \,, m)$
after Poisson summation, 
namely that given in
Eq.~\eqref{eq:Poissond}, 
we arrive at the second Taylor coefficient
\begin{eqnarray}
\mathcal{I}^{(2)}_\beta (m)
&=&
\frac{L^2}{4 ( 4 \pi)^{\frac{3}{2}} \Gamma(\b)}
\int_0^\infty ds \, s^{\b - \frac{5}{2}} e^{ - s m^2}
\notag\\
&& \times \vartheta''_3 \left(0,e^{- \frac{L^2}{4 s}} \right) \vartheta_3 \left(0,e^{- \frac{L^2}{4 s}} \right)^2
,\end{eqnarray}
which vanishes in infinite volume.

At this point, 
it is useful to recall the definition of the Jacobi elliptic-theta function
\begin{equation}
\vartheta_3 (z, q) = \sum_{\nu = -\infty}^\infty \cos ( 2 \nu z) q^{\nu^2}
,\end{equation}
and, 
as customary, 
primes denote derivatives with respect to the first argument, 
for example
$\vartheta'_3 (z, q) = \partial \vartheta_3 (z, q) / \partial z$. 
The elliptic-theta functions satisfy a diffusion equation
\begin{equation}
\left[
4 \frac{\partial}{\partial \eta} 
-
\frac{\partial^2}{\partial z^2} 
\right]
\vartheta_3 (z, e^{ - \eta}) 
= 
0
.\end{equation}
Consequently, 
we have the relation
\begin{eqnarray}
&&
\vartheta''_3 \left(0,e^{- \frac{L^2}{4 s}} \right) \vartheta_3 \left(0,e^{- \frac{L^2}{4 s}} \right)^2
=
\notag \\
&& \phantom{spacingsp}
-
\frac{16 s^2}{3  L^2} 
\frac{d}{ds}
\left[  
\vartheta_3 \left(0,e^{- \frac{L^2}{4 s}} \right)^3
-
1
\right]
.\end{eqnarray}
The arbitrary 
$s$-independent constant was added in order to avoid a surface term below. 
Returning to the expression for the second Taylor coefficient and performing an integration by parts, 
we see that
\begin{eqnarray}
\mathcal{I}^{(2)}_\beta (m)
&=&
\frac{4 \b (\b + 1)}{3 (4 \pi)^{\frac{3}{2}} \Gamma(\b +2)}
\int_0^\infty ds \,s^{ \beta - \frac{1}{2}} e^{ - s m^2}
\notag \\
&& 
\times
\left[ \frac{\b - \frac{1}{2}}{s}- m^2 \right]
\left[  
\vartheta_3 \left(0,e^{- \frac{L^2}{4 s}} \right)^3
-
1
\right]
.\notag \\
\end{eqnarray}
Comparing with Eq.~\eqref{eq:Poissond}, 
allows us to identify
\begin{eqnarray}
\mathcal{I}^{(2)}_\beta (m)
&=&
\frac{4}{3} \b (\beta +1)
\Bigg[
\frac{\b - \frac{1}{2}}{\beta + 1}
\mathcal{I}_{\b + 1} (m)
- 
m^2 \mathcal{I}_{\b + 2} (m)
\Bigg]
,\notag \\
\end{eqnarray}
which is precisely the Taylor series coefficient that we find from directly differentiating 
Eq.~\eqref{eq:I}
provided that the momentum 
$\vec{p} \,$
remains in the subtracted infinite volume integral. 
If this momentum is translated away, 
coefficients in the Taylor series expansion about zero momentum will no longer vanish in infinite volume.

\bibliography{zerobib}

\begin{thebibliography}{35}
\expandafter\ifx\csname natexlab\endcsname\relax\def\natexlab#1{#1}\fi
\expandafter\ifx\csname bibnamefont\endcsname\relax
  \def\bibnamefont#1{#1}\fi
\expandafter\ifx\csname bibfnamefont\endcsname\relax
  \def\bibfnamefont#1{#1}\fi
\expandafter\ifx\csname citenamefont\endcsname\relax
  \def\citenamefont#1{#1}\fi
\expandafter\ifx\csname url\endcsname\relax
  \def\url#1{\texttt{#1}}\fi
\expandafter\ifx\csname urlprefix\endcsname\relax\def\urlprefix{URL }\fi
\providecommand{\bibinfo}[2]{#2}
\providecommand{\eprint}[2][]{\url{#2}}

\bibitem[{\citenamefont{Pohl et~al.}(2010)\citenamefont{Pohl, Antognini, Nez,
  Amaro, Biraben et~al.}}]{Pohl:2010zza}
\bibinfo{author}{\bibfnamefont{R.}~\bibnamefont{Pohl}},
  \bibinfo{author}{\bibfnamefont{A.}~\bibnamefont{Antognini}},
  \bibinfo{author}{\bibfnamefont{F.}~\bibnamefont{Nez}},
  \bibinfo{author}{\bibfnamefont{F.~D.} \bibnamefont{Amaro}},
  \bibinfo{author}{\bibfnamefont{F.}~\bibnamefont{Biraben}},
  \bibnamefont{et~al.}, \bibinfo{journal}{Nature}
  \textbf{\bibinfo{volume}{466}}, \bibinfo{pages}{213} (\bibinfo{year}{2010}).

\bibitem[{\citenamefont{Antognini et~al.}(2013)\citenamefont{Antognini, Nez,
  Schuhmann, Amaro, Biraben et~al.}}]{Antognini:1900ns}
\bibinfo{author}{\bibfnamefont{A.}~\bibnamefont{Antognini}},
  \bibinfo{author}{\bibfnamefont{F.}~\bibnamefont{Nez}},
  \bibinfo{author}{\bibfnamefont{K.}~\bibnamefont{Schuhmann}},
  \bibinfo{author}{\bibfnamefont{F.~D.} \bibnamefont{Amaro}},
  \bibinfo{author}{\bibfnamefont{F.}~\bibnamefont{Biraben}},
  \bibnamefont{et~al.}, \bibinfo{journal}{Science}
  \textbf{\bibinfo{volume}{339}}, \bibinfo{pages}{417} (\bibinfo{year}{2013}).

\bibitem[{\citenamefont{Mohr et~al.}(2012)\citenamefont{Mohr, Taylor, and
  Newell}}]{Mohr:2012tt}
\bibinfo{author}{\bibfnamefont{P.~J.} \bibnamefont{Mohr}},
  \bibinfo{author}{\bibfnamefont{B.~N.} \bibnamefont{Taylor}},
  \bibnamefont{and} \bibinfo{author}{\bibfnamefont{D.~B.}
  \bibnamefont{Newell}}, \bibinfo{journal}{Rev.Mod.Phys.}
  \textbf{\bibinfo{volume}{84}}, \bibinfo{pages}{1527} (\bibinfo{year}{2012}),
  \eprint{1203.5425}.

\bibitem[{\citenamefont{Pohl et~al.}(2013)\citenamefont{Pohl, Gilman, Miller,
  and Pachucki}}]{Pohl:2013yb}
\bibinfo{author}{\bibfnamefont{R.}~\bibnamefont{Pohl}},
  \bibinfo{author}{\bibfnamefont{R.}~\bibnamefont{Gilman}},
  \bibinfo{author}{\bibfnamefont{G.~A.} \bibnamefont{Miller}},
  \bibnamefont{and} \bibinfo{author}{\bibfnamefont{K.}~\bibnamefont{Pachucki}},
  \bibinfo{journal}{Ann.Rev.Nucl.Part.Sci.} \textbf{\bibinfo{volume}{63}},
  \bibinfo{pages}{175} (\bibinfo{year}{2013}), \eprint{1301.0905}.

\bibitem[{\citenamefont{Syritsyn}(2014)}]{Syritsyn:2014saa}
\bibinfo{author}{\bibfnamefont{S.}~\bibnamefont{Syritsyn}},
  \bibinfo{journal}{PoS} \textbf{\bibinfo{volume}{LATTICE2013}},
  \bibinfo{pages}{009} (\bibinfo{year}{2014}), \eprint{1403.4686}.

\bibitem[{\citenamefont{Bedaque}(2004)}]{Bedaque:2004kc}
\bibinfo{author}{\bibfnamefont{P.~F.} \bibnamefont{Bedaque}},
  \bibinfo{journal}{Phys. Lett.} \textbf{\bibinfo{volume}{B593}},
  \bibinfo{pages}{82} (\bibinfo{year}{2004}), \eprint{nucl-th/0402051}.

\bibitem[{\citenamefont{de~Divitiis et~al.}(2004)\citenamefont{de~Divitiis,
  Petronzio, and Tantalo}}]{deDivitiis:2004kq}
\bibinfo{author}{\bibfnamefont{G.~M.} \bibnamefont{de~Divitiis}},
  \bibinfo{author}{\bibfnamefont{R.}~\bibnamefont{Petronzio}},
  \bibnamefont{and} \bibinfo{author}{\bibfnamefont{N.}~\bibnamefont{Tantalo}},
  \bibinfo{journal}{Phys. Lett.} \textbf{\bibinfo{volume}{B595}},
  \bibinfo{pages}{408} (\bibinfo{year}{2004}), \eprint{hep-lat/0405002}.

\bibitem[{\citenamefont{Sachrajda and Villadoro}(2005)}]{Sachrajda:2004mi}
\bibinfo{author}{\bibfnamefont{C.~T.} \bibnamefont{Sachrajda}}
  \bibnamefont{and}
  \bibinfo{author}{\bibfnamefont{G.}~\bibnamefont{Villadoro}},
  \bibinfo{journal}{Phys. Lett.} \textbf{\bibinfo{volume}{B609}},
  \bibinfo{pages}{73} (\bibinfo{year}{2005}), \eprint{hep-lat/0411033}.

\bibitem[{\citenamefont{Flynn et~al.}(2006)\citenamefont{Flynn, Juttner, and
  Sachrajda}}]{Flynn:2005in}
\bibinfo{author}{\bibfnamefont{J.}~\bibnamefont{Flynn}},
  \bibinfo{author}{\bibfnamefont{A.}~\bibnamefont{Juttner}}, \bibnamefont{and}
  \bibinfo{author}{\bibfnamefont{C.}~\bibnamefont{Sachrajda}}
  (\bibinfo{collaboration}{UKQCD Collaboration}), \bibinfo{journal}{Phys.Lett.}
  \textbf{\bibinfo{volume}{B632}}, \bibinfo{pages}{313} (\bibinfo{year}{2006}),
  \eprint{hep-lat/0506016}.

\bibitem[{\citenamefont{Guadagnoli et~al.}(2006)\citenamefont{Guadagnoli,
  Mescia, and Simula}}]{Guadagnoli:2005be}
\bibinfo{author}{\bibfnamefont{D.}~\bibnamefont{Guadagnoli}},
  \bibinfo{author}{\bibfnamefont{F.}~\bibnamefont{Mescia}}, \bibnamefont{and}
  \bibinfo{author}{\bibfnamefont{S.}~\bibnamefont{Simula}},
  \bibinfo{journal}{Phys.Rev.} \textbf{\bibinfo{volume}{D73}},
  \bibinfo{pages}{114504} (\bibinfo{year}{2006}), \eprint{hep-lat/0512020}.

\bibitem[{\citenamefont{Boyle et~al.}(2007)\citenamefont{Boyle, Flynn, Juttner,
  Sachrajda, and Zanotti}}]{Boyle:2007wg}
\bibinfo{author}{\bibfnamefont{P.}~\bibnamefont{Boyle}},
  \bibinfo{author}{\bibfnamefont{J.}~\bibnamefont{Flynn}},
  \bibinfo{author}{\bibfnamefont{A.}~\bibnamefont{Juttner}},
  \bibinfo{author}{\bibfnamefont{C.}~\bibnamefont{Sachrajda}},
  \bibnamefont{and} \bibinfo{author}{\bibfnamefont{J.}~\bibnamefont{Zanotti}},
  \bibinfo{journal}{JHEP} \textbf{\bibinfo{volume}{0705}}, \bibinfo{pages}{016}
  (\bibinfo{year}{2007}), \eprint{hep-lat/0703005}.

\bibitem[{\citenamefont{Boyle et~al.}(2008)\citenamefont{Boyle, Flynn, Juttner,
  Kelly, de~Lima et~al.}}]{Boyle:2008yd}
\bibinfo{author}{\bibfnamefont{P.}~\bibnamefont{Boyle}},
  \bibinfo{author}{\bibfnamefont{J.}~\bibnamefont{Flynn}},
  \bibinfo{author}{\bibfnamefont{A.}~\bibnamefont{Juttner}},
  \bibinfo{author}{\bibfnamefont{C.}~\bibnamefont{Kelly}},
  \bibinfo{author}{\bibfnamefont{H.~P.} \bibnamefont{de~Lima}},
  \bibnamefont{et~al.}, \bibinfo{journal}{JHEP}
  \textbf{\bibinfo{volume}{0807}}, \bibinfo{pages}{112} (\bibinfo{year}{2008}),
  \eprint{0804.3971}.

\bibitem[{\citenamefont{Frezzotti et~al.}(2009)\citenamefont{Frezzotti, Lubicz,
  and Simula}}]{Frezzotti:2008dr}
\bibinfo{author}{\bibfnamefont{R.}~\bibnamefont{Frezzotti}},
  \bibinfo{author}{\bibfnamefont{V.}~\bibnamefont{Lubicz}}, \bibnamefont{and}
  \bibinfo{author}{\bibfnamefont{S.}~\bibnamefont{Simula}}
  (\bibinfo{collaboration}{ETM Collaboration}), \bibinfo{journal}{Phys.Rev.}
  \textbf{\bibinfo{volume}{D79}}, \bibinfo{pages}{074506}
  (\bibinfo{year}{2009}), \eprint{0812.4042}.

\bibitem[{\citenamefont{Bedaque and Chen}(2005)}]{Bedaque:2004ax}
\bibinfo{author}{\bibfnamefont{P.~F.} \bibnamefont{Bedaque}} \bibnamefont{and}
  \bibinfo{author}{\bibfnamefont{J.-W.} \bibnamefont{Chen}},
  \bibinfo{journal}{Phys.Lett.} \textbf{\bibinfo{volume}{B616}},
  \bibinfo{pages}{208} (\bibinfo{year}{2005}), \eprint{hep-lat/0412023}.

\bibitem[{\citenamefont{B{\"a}r et~al.}(2003)\citenamefont{B{\"a}r, Rupak, and
  Shoresh}}]{Bar:2002nr}
\bibinfo{author}{\bibfnamefont{O.}~\bibnamefont{B{\"a}r}},
  \bibinfo{author}{\bibfnamefont{G.}~\bibnamefont{Rupak}}, \bibnamefont{and}
  \bibinfo{author}{\bibfnamefont{N.}~\bibnamefont{Shoresh}},
  \bibinfo{journal}{Phys. Rev.} \textbf{\bibinfo{volume}{D67}},
  \bibinfo{pages}{114505} (\bibinfo{year}{2003}), \eprint{hep-lat/0210050}.

\bibitem[{\citenamefont{de~Divitiis et~al.}(2012)\citenamefont{de~Divitiis,
  Petronzio, and Tantalo}}]{deDivitiis:2012vs}
\bibinfo{author}{\bibfnamefont{G.}~\bibnamefont{de~Divitiis}},
  \bibinfo{author}{\bibfnamefont{R.}~\bibnamefont{Petronzio}},
  \bibnamefont{and} \bibinfo{author}{\bibfnamefont{N.}~\bibnamefont{Tantalo}},
  \bibinfo{journal}{Phys.Lett.} \textbf{\bibinfo{volume}{B718}},
  \bibinfo{pages}{589} (\bibinfo{year}{2012}), \eprint{1208.5914}.

\bibitem[{\citenamefont{Tiburzi}(2008)}]{Tiburzi:2007ep}
\bibinfo{author}{\bibfnamefont{B.~C.} \bibnamefont{Tiburzi}},
  \bibinfo{journal}{Phys.Rev.} \textbf{\bibinfo{volume}{D77}},
  \bibinfo{pages}{014510} (\bibinfo{year}{2008}), \eprint{0710.3577}.

\bibitem[{\citenamefont{Burkardt}(2000)}]{Burkardt:2000za}
\bibinfo{author}{\bibfnamefont{M.}~\bibnamefont{Burkardt}},
  \bibinfo{journal}{Phys.Rev.} \textbf{\bibinfo{volume}{D62}},
  \bibinfo{pages}{071503} (\bibinfo{year}{2000}), \eprint{hep-ph/0005108}.

\bibitem[{\citenamefont{Burkardt}(2003)}]{Burkardt:2002hr}
\bibinfo{author}{\bibfnamefont{M.}~\bibnamefont{Burkardt}},
  \bibinfo{journal}{Int.J.Mod.Phys.} \textbf{\bibinfo{volume}{A18}},
  \bibinfo{pages}{173} (\bibinfo{year}{2003}), \eprint{hep-ph/0207047}.

\bibitem[{\citenamefont{Draper et~al.}(1989)\citenamefont{Draper, Woloshyn,
  Wilcox, and Liu}}]{Draper:1988bp}
\bibinfo{author}{\bibfnamefont{T.}~\bibnamefont{Draper}},
  \bibinfo{author}{\bibfnamefont{R.}~\bibnamefont{Woloshyn}},
  \bibinfo{author}{\bibfnamefont{W.}~\bibnamefont{Wilcox}}, \bibnamefont{and}
  \bibinfo{author}{\bibfnamefont{K.-F.} \bibnamefont{Liu}},
  \bibinfo{journal}{Nucl.Phys.} \textbf{\bibinfo{volume}{B318}},
  \bibinfo{pages}{319} (\bibinfo{year}{1989}).

\bibitem[{\citenamefont{Bunton et~al.}(2006)\citenamefont{Bunton, Jiang, and
  Tiburzi}}]{Bunton:2006va}
\bibinfo{author}{\bibfnamefont{T.}~\bibnamefont{Bunton}},
  \bibinfo{author}{\bibfnamefont{F.-J.} \bibnamefont{Jiang}}, \bibnamefont{and}
  \bibinfo{author}{\bibfnamefont{B.}~\bibnamefont{Tiburzi}},
  \bibinfo{journal}{Phys.Rev.} \textbf{\bibinfo{volume}{D74}},
  \bibinfo{pages}{034514} (\bibinfo{year}{2006}), \eprint{hep-lat/0607001}.

\bibitem[{\citenamefont{Aubin et~al.}(2013)\citenamefont{Aubin, Blum,
  Golterman, and Peris}}]{Aubin:2013daa}
\bibinfo{author}{\bibfnamefont{C.}~\bibnamefont{Aubin}},
  \bibinfo{author}{\bibfnamefont{T.}~\bibnamefont{Blum}},
  \bibinfo{author}{\bibfnamefont{M.}~\bibnamefont{Golterman}},
  \bibnamefont{and} \bibinfo{author}{\bibfnamefont{S.}~\bibnamefont{Peris}},
  \bibinfo{journal}{Phys.Rev.} \textbf{\bibinfo{volume}{D88}},
  \bibinfo{pages}{074505} (\bibinfo{year}{2013}), \eprint{1307.4701}.

\bibitem[{\citenamefont{Horch et~al.}(2013)\citenamefont{Horch, Herdo{\'i}za,
  J{\"a}ger, Wittig, Della~Morte et~al.}}]{Horch:2013lla}
\bibinfo{author}{\bibfnamefont{H.}~\bibnamefont{Horch}},
  \bibinfo{author}{\bibfnamefont{G.}~\bibnamefont{Herdo{\'i}za}},
  \bibinfo{author}{\bibfnamefont{B.}~\bibnamefont{J{\"a}ger}},
  \bibinfo{author}{\bibfnamefont{H.}~\bibnamefont{Wittig}},
  \bibinfo{author}{\bibfnamefont{M.}~\bibnamefont{Della~Morte}},
  \bibnamefont{et~al.}, \bibinfo{journal}{PoS}
  \textbf{\bibinfo{volume}{LATTICE2013}}, \bibinfo{pages}{304}
  (\bibinfo{year}{2013}), \eprint{1311.6975}.

\bibitem[{\citenamefont{Bernard and Golterman}(1994)}]{Bernard:1993sv}
\bibinfo{author}{\bibfnamefont{C.~W.} \bibnamefont{Bernard}} \bibnamefont{and}
  \bibinfo{author}{\bibfnamefont{M.~F.} \bibnamefont{Golterman}},
  \bibinfo{journal}{Phys.Rev.} \textbf{\bibinfo{volume}{D49}},
  \bibinfo{pages}{486} (\bibinfo{year}{1994}), \eprint{hep-lat/9306005}.

\bibitem[{\citenamefont{Sharpe and Shoresh}(2000)}]{Sharpe:2000bc}
\bibinfo{author}{\bibfnamefont{S.~R.} \bibnamefont{Sharpe}} \bibnamefont{and}
  \bibinfo{author}{\bibfnamefont{N.}~\bibnamefont{Shoresh}},
  \bibinfo{journal}{Phys.Rev.} \textbf{\bibinfo{volume}{D62}},
  \bibinfo{pages}{094503} (\bibinfo{year}{2000}), \eprint{hep-lat/0006017}.

\bibitem[{\citenamefont{Sharpe and Shoresh}(2001)}]{Sharpe:2001fh}
\bibinfo{author}{\bibfnamefont{S.~R.} \bibnamefont{Sharpe}} \bibnamefont{and}
  \bibinfo{author}{\bibfnamefont{N.}~\bibnamefont{Shoresh}},
  \bibinfo{journal}{Phys.Rev.} \textbf{\bibinfo{volume}{D64}},
  \bibinfo{pages}{114510} (\bibinfo{year}{2001}), \eprint{hep-lat/0108003}.

\bibitem[{\citenamefont{Gasser and Leutwyler}(1988)}]{Gasser:1987zq}
\bibinfo{author}{\bibfnamefont{J.}~\bibnamefont{Gasser}} \bibnamefont{and}
  \bibinfo{author}{\bibfnamefont{H.}~\bibnamefont{Leutwyler}},
  \bibinfo{journal}{Nucl.Phys.} \textbf{\bibinfo{volume}{B307}},
  \bibinfo{pages}{763} (\bibinfo{year}{1988}).

\bibitem[{\citenamefont{Arndt and Tiburzi}(2003)}]{Arndt:2003ww}
\bibinfo{author}{\bibfnamefont{D.}~\bibnamefont{Arndt}} \bibnamefont{and}
  \bibinfo{author}{\bibfnamefont{B.~C.} \bibnamefont{Tiburzi}},
  \bibinfo{journal}{Phys. Rev.} \textbf{\bibinfo{volume}{D68}},
  \bibinfo{pages}{094501} (\bibinfo{year}{2003}), \eprint{hep-lat/0307003}.

\bibitem[{\citenamefont{Bijnens and Relefors}(2014)}]{Bijnens:2014yya}
\bibinfo{author}{\bibfnamefont{J.}~\bibnamefont{Bijnens}} \bibnamefont{and}
  \bibinfo{author}{\bibfnamefont{J.}~\bibnamefont{Relefors}},
  \bibinfo{journal}{JHEP} \textbf{\bibinfo{volume}{1405}}, \bibinfo{pages}{015}
  (\bibinfo{year}{2014}), \eprint{1402.1385}.

\bibitem[{\citenamefont{Jiang and Tiburzi}(2008{\natexlab{a}})}]{Jiang:2008ja}
\bibinfo{author}{\bibfnamefont{F.-J.} \bibnamefont{Jiang}} \bibnamefont{and}
  \bibinfo{author}{\bibfnamefont{B.}~\bibnamefont{Tiburzi}},
  \bibinfo{journal}{Phys.Rev.} \textbf{\bibinfo{volume}{D78}},
  \bibinfo{pages}{114505} (\bibinfo{year}{2008}{\natexlab{a}}),
  \eprint{0810.1495}.

\bibitem[{\citenamefont{Hu et~al.}(2007)\citenamefont{Hu, Jiang, and
  Tiburzi}}]{Hu:2007eb}
\bibinfo{author}{\bibfnamefont{J.}~\bibnamefont{Hu}},
  \bibinfo{author}{\bibfnamefont{F.-J.} \bibnamefont{Jiang}}, \bibnamefont{and}
  \bibinfo{author}{\bibfnamefont{B.~C.} \bibnamefont{Tiburzi}},
  \bibinfo{journal}{Phys.Lett.} \textbf{\bibinfo{volume}{B653}},
  \bibinfo{pages}{350} (\bibinfo{year}{2007}), \eprint{0706.3408}.

\bibitem[{\citenamefont{Jiang and Tiburzi}(2007)}]{Jiang:2006gna}
\bibinfo{author}{\bibfnamefont{F.-J.} \bibnamefont{Jiang}} \bibnamefont{and}
  \bibinfo{author}{\bibfnamefont{B.}~\bibnamefont{Tiburzi}},
  \bibinfo{journal}{Phys.Lett.} \textbf{\bibinfo{volume}{B645}},
  \bibinfo{pages}{314} (\bibinfo{year}{2007}), \eprint{hep-lat/0610103}.

\bibitem[{\citenamefont{Jiang and Tiburzi}(2008{\natexlab{b}})}]{Jiang:2008te}
\bibinfo{author}{\bibfnamefont{F.-J.} \bibnamefont{Jiang}} \bibnamefont{and}
  \bibinfo{author}{\bibfnamefont{B.~C.} \bibnamefont{Tiburzi}},
  \bibinfo{journal}{Phys.Rev.} \textbf{\bibinfo{volume}{D78}},
  \bibinfo{pages}{037501} (\bibinfo{year}{2008}{\natexlab{b}}),
  \eprint{0806.4371}.

\bibitem[{\citenamefont{Beringer et~al.}(2012)}]{Beringer:1900zz}
\bibinfo{author}{\bibfnamefont{J.}~\bibnamefont{Beringer}} \bibnamefont{et~al.}
  (\bibinfo{collaboration}{Particle Data Group}), \bibinfo{journal}{Phys.Rev.}
  \textbf{\bibinfo{volume}{D86}}, \bibinfo{pages}{010001}
  (\bibinfo{year}{2012}).

\bibitem[{\citenamefont{Gasser and Leutwyler}(1985)}]{Gasser:1985ux}
\bibinfo{author}{\bibfnamefont{J.}~\bibnamefont{Gasser}} \bibnamefont{and}
  \bibinfo{author}{\bibfnamefont{H.}~\bibnamefont{Leutwyler}},
  \bibinfo{journal}{Nucl. Phys.} \textbf{\bibinfo{volume}{B250}},
  \bibinfo{pages}{517} (\bibinfo{year}{1985}).

\end{thebibliography}

\end{document}